# Using realistic host galaxy metallicities to improve the GRB X-ray equivalent total hydrogen column density and constrain the intergalactic medium density

Tony Dalton[1] and Simon L. Morris[1]

[1] Centre for Extragalactic Astronomy, Durham University, South Road, Durham DH1 3LE, UK

E-mail: tonydalton@live.ie

**ABSTRACT**

It is known that the GRB equivalent hydrogen column density ($N_{HX}$) changes with redshift and that, typically, $N_{HX}$ is greater than the GRB host neutral hydrogen column density. We have compiled a large sample of data for GRB $N_{HX}$ and metallicity [X/H]. The main aims of this paper are to generate improved $N_{HX}$ for our sample by using actual metallicities, dust corrected where available for detections, and for the remaining GRB, a more realistic average intrinsic metallicity using a standard adjustment from solar. Then, by approximating the GRB host intrinsic hydrogen column density using the measured neutral column ($N_{HI,IC}$) adjusted for the ionisation fraction, we isolate a more accurate estimate for the intergalactic medium (IGM) contribution. The GRB sample mean metallicity is = -1.17±0.09 rms (or 0.07±0.05 $Z/Z_{sol}$) from a sample of 36 GRB with a redshift $1.76 \leq z \leq 5.91$, substantially lower than the assumption of solar metallicity used as standard for many fitted $N_{HX}$. Lower GRB host mean metallicity results in increased estimated $N_{HX}$ with the correction scaling with redshift as $\Delta \log (N_{HX}$ cm$^{-2}$) = (0.59±0.04)log(1+z) + 0.18±0.02. Of the 128 GRB with data for both $N_{HX}$ and $N_{HI,IC}$ in our sample, only 6 have $N_{HI,IC} > N_{HX}$ when revised for realistic metallicity, compared to 32 when solar metallicity is assumed. The lower envelope of the revised $N_{HX}$ - $N_{HI,IC}$, plotted against redshift can be fit by log($N_{HX}$ - $N_{HI,IC}$ cm$^{-2}$) =20.3 + 2.4 log(1+z). This is taken to be an estimate for the maximum IGM hydrogen column density as a function of redshift. Using this approach, we estimate an upper limit to the hydrogen density at redshift zero ($n_0$) to be consistent with $n_0$ = 0.17 x $10^{-7}$cm$^{-3}$.

## 1 INTRODUCTION

Gamma-ray bursts (GRBs) are among the most powerful explosions known in the universe, (see Schady (2017, S17 hereafter) for a recent general review of GRB). Given the huge range of redshifts and distances for GRBs, and their high luminosities combined with the broad energy range of observed emissions, GRBs provide a valuable probe of all baryonic matter along the line-of-sight. X-ray absorption yields information on the total absorbing column density of the matter between the observer and the source because any element that is not fully ionized contributes to the absorption of X-rays (scattering by electrons becomes important at high energy above 10keV (Wilms, Allen and McCray, 2000, hereafter W00) . Although the X-ray absorption cross-section is often dominated by metals, with hydrogen and helium contribution being minimal but not nil (Fig.1 W00), it is typically reported as an equivalent hydrogen column density (in this paper $N_{HX}$ ). $N_{HX}$ consists of contributions from the local GRB environment, the intergalactic medium (IGM), and our own Galactic medium. However, X-ray absorption cannot reveal the redshift of the matter in the column due to a lack of spectral resolution and signal to noise. It is important to note that the common practice is to make the simplifying assumption that all X-ray absorption in excess of Galactic is at the redshift of the host, neglecting any IGM contribution (e.g. Watson et al., 2007; Starling et al., 2013).The GRB $N_{HX}$ versus redshift relation has been investigated for many years. Early reports were based on small samples (e.g. Campana et al., 2010; Behar et al., 2011; Watson et al., 2013). A claimed strong correlation with redshift has recently been updated and confirmed with a much larger GRB sample by Rahin and Behar, (2019). It has also been reported in many papers that the neutral intrinsic hydrogen column ($N_{HI}$) in GRB has no significant correlation with redshift (e.g. Watson et al., 2007). Further, it was also noted in these papers, that $N_{HX}$ exceeds $N_{HI}$ in GRB, often by over an order of magnitude.

The cause of an $N_{HX}$ excess over $N_{HI}$, and the $N_{HX}$ correlation with redshift is the source of much debate. One school of thought argues that the GRB host accounts for all the excess and evolution, e.g. a dense environment near the burst location (Campana et al., 2012, hereafter C12), ultra-ionised gas in the environment of the GRB (Schady et al., 2011, hereafter S11), dust extinction bias (Watson and Jakobsson, 2012), dense Helium (He II) regions close to the GRB (Watson et al., 2013), and/or a host galaxy mass $N_{HX}$ relation (Buchner, Schulze and Bauer, 2017). Models for GRB $N_{HX}$ being produced exclusively by gas intrinsic to the GRB host galaxy have required extreme conditions to be present within the absorbing material. The other school of thought argues that some of the excess $N_{HX}$ and redshift correlation is due to the full integrated line of sight (LOS) including the diffuse IGM and intervening objects. Behar et al., (2011) modelled the effects of a cold, neutral, highly metal-enriched IGM model and showed that, at high redshift, this could produce the dominant excess X-ray absorption component. Starling et al., (2013, S13 hereafter) modelled a more realistic warm IGM (WHIM) with temperature of between $10^5$-$10^{6.5}$ K and metallicity of ~0.2 $Z/Z_{sol}$. Campana et al., (2015) used cosmological simulations to model the WHIM. Their results suggested that most of the excess $N_{HX}$ absorption arises from discrete over-densities along the line of sight (LOS) to GRB, supporting the possibility of a significant contribution of the IGM to the $N_{HX}$ - redshift relation.

All of the theories thus far have relied upon key assumptions (listed below) which, if unrealistic, will substantially affect the results.

### 1.1 Metallicity

It is known that GRB galaxy hosts have, on average, sub-solar metallicity, and that assuming solar metallicity in the X-ray fits introduces a systematic error, and generates an $N_{HX}$ that is effectively, a minimum (S13; Krühler et al., 2015; Tanga et al., 2016). Further, models for the WHIM integrated gas density along the LOS heavily rely on the assumed gas metallicity of the WHIM. It is standard practice currently to assume solar metallicity when fitting models to GRB X-ray spectra. The main reasons for this historically were the small numbers of reliable metallicity measurements and poor constraints on any redshift metallicity evolution (S13). Even an assumption of solar metallicity, however, can lead to inconsistencies, as research improves the knowledge of solar abundances. The solar abundances of the key metals reported in the literature have undergone considerable changes in recent decades (for a useful review see Asplund et al., (2009)). The X-ray fitting software XSPEC[1] (Arnaud, 1996) is the most commonly used for GRB. Within XSPEC, the default solar abundances are those of Anders and Grevesse, (1989). However, there are six other options for solar abundance in XSPEC. The more commonly used is that of W00. S13 noted that their results using W00 abundances were consistently higher than the $N_{HX}$ reported in the UK Swift Science Data Centre[2] repository (hereafter *Swift*), which at the time were based on Anders and Grevesse, (1989). These have since been updated using W00.

Some comments have been made in literature as to how $N_{HX}$ scales with metallicity e.g. S13 stated that $N_{HX}$ scales approximately with metallicity, and Krongold and Prochaska, (2013) stated that the X-ray estimated oxygen column density has a linear dependence on metallicity and density. Metallicity is the main focus of this paper and Section 3 examines the impact of the assumed metallicity on the derived $N_{HX}$ in some detail.

### 1.2 Location of excess absorption

It is standard practice, when fitting models to GRB spectra, to assume all absorption in excess of the Galactic contribution is at the redshift of the GRB. X-ray optical depth is a function of frequency or energy, due to the frequency dependence of the cross-section (Morrison & Mccammon, 1983). The scaling relation between the observed amount of X-ray absorption for GRB and redshift was found by Campana et al., (2014) to be approximated by

$$N_H (z = 0) = N_H(z)/(1 + z)^a, a = 2.4 \quad (1)$$

The error in X-ray column density produced by assuming the total absorption is at the GRB redshift arises from the difference in redshift between the GRB and any intervening contributor. Hence, the potential error in $N_{HX}$ increases with redshift of the GRB, dependant on the amount of IGM absorption, its location and any error in the scaling law assumed. The IGM hydrogen column estimation is highly uncertain as the metal pollution is very poorly determined (e.g. Fumagalli, 2014; Maiolino & Mannucci, 2019).

### 1.3 Neutral fraction

The value found for the column density is almost always determined assuming a 100% neutral absorbing gas (e.g. Behar et al., 2011; S11; S13). An ionized absorber would have a lower cross-section at X-ray energies, and a larger column density would be required to produce the same opacity.

---

[1] https://heasarc.gsfc.nasa.gov/xanadu/xspec/

[2] www.swift.ac.uk/xrt_spectra

Therefore, the neutral assumption could cause $N_{HX}$ to be underestimated (S11).

*1.4 Galactic absorption*

Column densities for GRB reported have normally had the Galaxy contribution ($N_{HGal}$) removed. The most common references for the $N_{HGal}$ are the Leiden Argentine Bonn (LAB) HI survey (Kalberla et al., 2005) and Willingale et al., (2013).

In conclusion, $N_{HX}$, based on an assumed solar metallicity and 100% neutral absorbing gas should be considered as a lower limit. Further, the inconsistent use or lack of reporting of the assumed metallicity, neutral fraction and scaling factors can add uncertainty to any analysis using published data.

The hypothesis of this paper is that the IGM contributes to the total hydrogen column density, with the contribution increasing with redshift, as observed in GRB $N_{HX}$, and that by correcting the GRB $N_{HX}$ using a more realistic GRB intrinsic metallicity, and estimating the host $N_H$ using the measured neutral intrinsic $N_{HI}$ adjusted for ionisation fraction (from optical spectra (rest-frame UV) of GRB afterglow), we can isolate a more accurate $N_{HIGM}$ contribution.

The objectives of this paper are:
- A review of the literature on the metallicities of GRB host environments to obtain improved values to use when estimating $N_{HX}$,
- To present a revised GRB $N_{HX}$, using these more realistic metallicities and hence to update the $N_{HX}$ - redshift relation.
- To isolate the IGM contribution to the total $N_{HX}$ in GRB, by using GRB ionised corrected $N_{HI}$ ($N_{HI,IC}$) as an estimate of the GRB intrinsic $N_{HX}$ and plotting $N_{HX}$ - $N_{HI,IC}$ against redshift, after the improved metal corrections have been used.
- To compare an estimated $N_{HIGM}$ based on a simple model of the IGM with our lower envelope for $N_{HX}$ based on realistic metallicities and with the intrinsic $N_{HI,IC}$ removed.

Section 2 sets out the methodology, data selection approach and the data used. Section 3 presents the results with a discussion and an analysis. Section 4 sets out the main conclusions. The Appendix gives further details on the Section 3 analysis, including the metallicity and the resulting fractional increase in $N_{HX}$ with redshift. Throughout this paper, the term "metallicity" is used synonymously with metal abundance [X/H][3]. Where relevant, the ΛCDM cosmology variables used are $H_0$ = 70 km s$^{-1}$ Mpc$^{-1}$, $\Omega_m$ = 0.3, and $\Omega_\Lambda$ = 0.7 unless otherwise stated.

---

[3] [X/H] = log(X/H)-log(X/H)$_{solar}$, where X is the metal element, and H is Hydrogen

## 2 METHODOLOGY AND DATA SELECTION

The full sample used here consists of all *Swift* X-ray Telescope (XRT) (Burrows et al., 2005) observed GRB with spectroscopic redshift available up to 31 July 2019, plus GRB090429B which has a photometric redshift of 9.4 (Cucchiara et al., 2011). The vast majority of the $N_{HX}$ data is taken from the *Swift* repository to ensure a homogeneous dataset (S13). Alternative sources were used only where detections with measured errors were available, and where *Swift* has only reported column density lower limits consistent with zero, or where the errors reported in the alternative source were smaller (e.g. Arcodia, Campana and Salvaterra, (2016) used a stricter methodology by selecting specific time intervals when hardness ratios were constant to minimise spectral variations). For all sources, we endeavoured to ensure the methodology and selection criteria were consistent in terms of confidence level and XSPEC models used. Data from the *Swift* repository for $N_{HX}$ were taken from the Photon Counting Late Time mode, as they are most likely to be a more stable, final value since spectral slope evolution is more prevalent at early times, leading to poor quality fits using a single power law (Page, K., private communication). All $N_{HX}$ error bars reported are the 90% confidence range, unless otherwise stated.

We follow the *Swift* repository reporting conventions for $N_{HX}$ i.e. we treat as an upper limit the cases where the best fit $N_{HX}$ is zero. Further, where the lower limit of the 90% confidence interval includes zero, we use the best fit $N_{HX}$ but use a different symbol for these objects in our figures.

Where we refitted spectra for analysis, XSPEC v12.10.1 was used (Arnaud, 1996). Spectra were fitted with a power law in the X-ray band from 0.3 – 10.0 keV, which is suitable for the vast majority of GRB and again is consistent with the *Swift* repository (S13). A fixed Galactic component is taken from *Swift* based on Willingale et al., (2013). The model used in XSPEC was *tbabs\*ztbabs\*po* where the initial assumption we want to use is that all absorption in excess of Galactic is at the host redshift. *tbabs* is the galactic ISM absorption model, *ztbabs* is the same model placing the absorption at a fixed redshift and *po* is the power law intrinsic spectral model. Isotopic abundances from W00 were used with the assumption of solar metallicity initially. In Section 3, where we examine more realistic metallicities for GRBs, the XSPEC model *tbvarabs* was used instead of *ztbabs* which allows the individual metal abundances to be varied from solar values. Cash statistics (Cstat) were used in XSPEC as this is required for spectra with low count rates, and is consistent with the *Swift* repository (Cash, 1979). In all refits, we took the best fits based on minimum reduced χ-squared.

**Table 1.** The GRB full sample. Column 1) GRB identification, 2) spectroscopic redshift (photometric for GRB090429B), 3) log($N_{HX}$ cm$^{-2}$), 4) Refs for log($N_{HX}$ cm$^{-2}$)( note all are from the *Swift* repository if no ref given), 5) log ($N_{HI}$ cm$^{-2}$) (all from (Tanvir et al., 2019). Those with "IC" have been corrected for ionisation fraction, 6) [X/H], 7) refs for [X/H].

| GRB | z | log($N_{HX}$ cm$^{-2}$) | $N_{HX}$ Ref. | log($N_{HI,IC}$ cm$^{-2}$) | [X/H] | [X/H] Ref. |
|---|---|---|---|---|---|---|
| 090926A | 2.11 | $21.74^{+0.19}_{-0.17}$ | 1 | 21.55±0.10 | -1.97±0.11 | 2 |
| 090809 | 2.74 | $21.85^{+0.27}_{-0.85}$ | | 21.70±0.20 | -0.86±0.13 | 2 |
| 080210 | 2.64 | $22.32^{+0.21}_{-0.32}$ | | 21.90±0.10 | -1.21±0.16 | 3 |
| 090313 | 3.38 | $22.64^{+0.13}_{-0.18}$ | | 21.30±0.20 | -1.40±0.30 | 3 |
| 120909A | 3.93 | $22.41^{+0.16}_{-0.24}$ | | 21.70±010 | -1.06±0.20 | 2 |

Table 1 Refs: (1) (Zafar et al., 2018), (2) (Bolmer et al., 2019), (3) (Arabsalmani et al., 2018)

The selection of a sample can introduce bias. Perley et al., (2016) found that GRB with measured redshifts tend to be found in brighter galaxies, which could produce such a bias. However, Rahin and Behar, (2019) compared the $N_{HX}$ versus redshift trend for that paper's (smaller) unbiased sample with the full *Swift* sample and found no notable difference. In, Section 3, where we require that GRB have both optical and X-ray spectra, this requirement can introduce a selection effect against dim or highly dust extinguished GRBs (S11). It is estimated that between 25-40% of GRBs are undetected in the optical wavelength range as a result of dust extinction (e.g. Greiner et al., 2011). Therefore, the conclusions in that section may not apply to dust extinguished or dim GRBs.

All data for $N_{HI}$ were taken from Tanvir et al., (2019). Of the 140 objects in their sample, we have used 128 which have $N_{HX}$ data in our analysis. In Section 3, we adjust the $N_{HI}$ for the ionisation fraction for log($N_{HI}$ cm$^{-2}$) < 20. The redshift range in our full GRB $N_{HX}$ sample is from 0.03 to 9.4, while the $N_{HI}$ sample range is from 1.6 to 6.73. The lower $N_{HI}$ redshift cut-off is due to the requirement that the observed wavelength of the Lyα absorption line be in the visible/UV band.

In Section 3.5, we analyse the impacts on $N_{HX}$ of using metallicities that more realistically reflect the LOS absorption through the entire host galaxy. X-ray absorption is dominated by the metals and H and He are relatively unimportant. Below 1 keV, C, N, O, and Ne are the main absorbers, while above 1 keV, Si, S, and Fe dominate (W00). W00 also note that interpreting X-ray observations is subject to the uncertainties remaining in the atomic data. Data for metallicities in Table 1 are all UV/optical absorption line based. Absorption metallicities measure the metal enrichment of gas along the LOS from the GRB through the galaxy.

Table 1 contains the data for the full GRB sample for redshift, $N_{HX}$, $N_{HI,IC}$ and metallicity where data is available. We list in the table extract, a sample of 5 GRB with data for all columns (see the online version of this paper for the complete table with all values listed). (See Appendix A1 an investigation about whether a flux limited sample would introduce any substantial bias).

## 3 RESULTS AND ANALYSIS

In this section, we examine the distributions of GRB $N_{HX}$ and $N_{HI}$ with redshift, and the use of adjusted $N_{HI,IC}$ as an approximation for the GRB host intrinsic contribution to the total integrated hydrogen column density to isolate the IGM column density. We then examine GRB host metallicity to derive a more accurate metallicity to use in XSPEC fitting to get an improved $N_{HX}$. Finally, we replot the distributions of revised $N_{HX}$ and $N_{HX}$ - $N_{HI,IC}$ with redshift to get the lower envelope of GRB $N_{HX}$ as a step towards constraining $N_{HIGM}$.

### 3.1 $N_{HX}$ AND REDSHIFT

Figure 1 shows the distribution of $N_{HX}$ with redshift (based on the standard assumptions of solar metallicity and that all the absorption, $N_{HX}$ is at the GRB redshift) for the full *Swift* observed sample with known spectroscopic redshift (with the exception of GRB090429B where the redshift is a photometric estimate). Where an estimate of the actual $N_{HX}$ was available from the *Swift* repository but the 90% confidence interval included zero, these are plotted with a yellow dot. Yellow dots with arrows are the upper limits where the *Swift* repository has a best fit of zero $N_{HX}$. A relationship or dependence between $N_{HX}$ and redshift has been reported in several papers over the last decade (e.g. Behar et al., 2011; S13). The correlation statistics for the full 352 GRB sample are Pearson r=0.29 and Spearman ρ= 0.55. For the detection only sample (226 GRB), the correlation results are Pearson r= 0.51 and Spearman ρ=0.49. Both samples pass the null hypothesis test, indicating that the correlation seen is significant. However, when we used an error weighted least squares fit to a linear model ($\chi^2$), the reduced $\chi^2$ was large, indicating either that a simple linear redshift relation is not a realistic model or that there is an additional substantial source of scatter. We checked for the impact of outliers with high $N_{HX}$ and very small error bars. To bring the reduced $\chi^2$ close to 1 would have required the removal of over 10% of the sample.

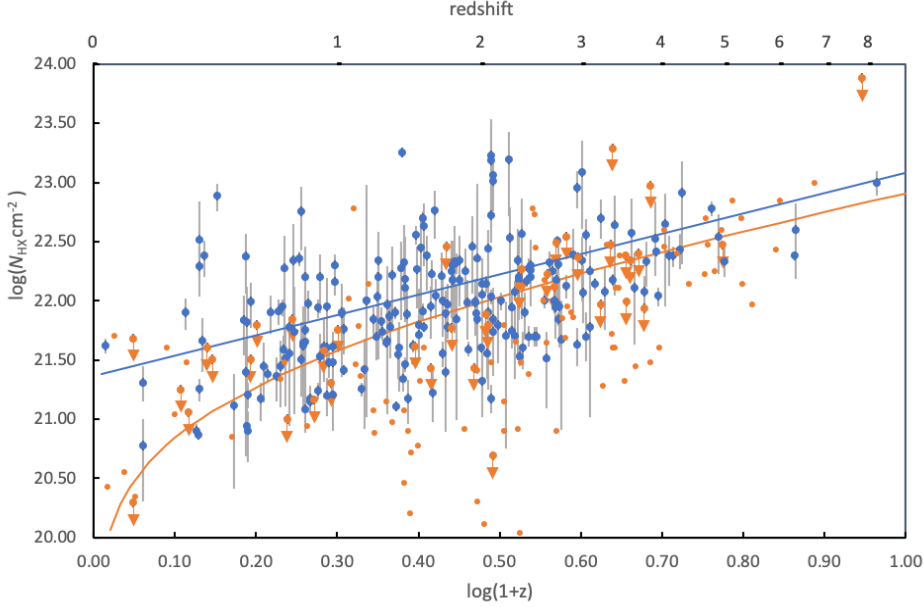

**Figure 1.** Distribution of intrinsic X-ray column densities ($N_{HX}$ cm$^{-2}$) with redshift for the full 352 GRB *Swift* observed sample. The blue dots represent the GRB detections with error bars. The orange dots are best fits where the 90% confidence interval includes zero. Orange dots with arrows are the upper limits where *Swift* repository has a best fit of zero. The blue line is the $\chi^2$ best fit with for the GRB data with error bars. The orange line represents the integrated hydrogen density $N_{HIGM}$ cm$^{-2}$) from a simple diffuse IGM model (see equation (2)). The correlation statistics for the full 352 GRB sample are Pearson r=0.29 and Spearman ρ = 0.55 (for the detection only sample (226 GRB) r=0.51 and ρ=0.49).

We include in Figure 1 a simple model of the diffuse IGM following S13 (equation 5 in that paper) based on

$$N_{HIGM} = (n_0 c/ H_0) \int_0^z (1+z')^2/((1+z')^3 \Omega_M + \Omega_\Lambda)^{1/2} \, dz' \quad (2)$$

where $n_0$ is the hydrogen density at redshift zero, taken as $1.7 \times 10^{-7}$ cm$^{-3}$ from Behar et al., (2011).

The solution to the integral from Shull and Danforth, (2018) (equation 4 in that paper) is

$$(2/(3\,\Omega_M))\{(\Omega_M(1+z)^3 + \Omega_\Lambda)^{1/2} - 1\} \quad (3)$$

In Figure 1, we can see the $N_{HIGM}$ model runs through the GRB datapoints. If it is to represent the diffuse IGM only, we would expect all the GRB to be above this curve, (if there were no measurement errors). Given the large error bars for many GRB, the IGM hypothesis could still be plausible where a small fraction, 10% approximately given the 90% confidence, are below the curve. A much higher fraction than 10% are below the IGM curve in Figure 1. However, the IGM model is admittedly very simple and therefore could poorly represent the real universe. Also, not all LOS will be at the mean density.

We note that our model is based on the mean hydrogen density as a simple model, so the metallicity uncertainty in the IGM does not affect it directly. In our next paper, the metallicity in different phases of the IGM will be reviewed in detail. We examine this further in Section 3.3 onwards.

### 3.2 $N_{HI}$ REVIEW WITH REDSHIFT

In this section, we review the most recent substantial GRB $N_{HI}$ sample from Tanvir et al., (2019) which consists of new measurements combined with those from literature. We examine this latest sample for any relations between $N_{HI}$ and redshift, or with $N_{HX}$.

Optical spectroscopy enables the approximate location of any neutral hydrogen absorber to be identified. GRB hosts are typically found to have high column densities of cold neutral gas, with a large fraction of GRB hosts containing a DLA system (log($N_{HI}$ cm$^{-2}$)> 20.3) or sub-DLA (19.0 < log($N_{HI}$ cm$^{-2}$)< 20.3) (S11). Much of the neutral gas component is found at a few hundred parsecs from the GRB (Ledoux et al., 2009).

Figure 2 shows the distribution of the GRB $N_{HI}$ sample with redshift. Where not specified in literature, we have assumed the errors are gaussian and correspond to one standard deviation. The Pearson r is -0.15 and Spearman ρ -0.10. Both fail the null hypothesis test i.e. there is no statistically significant correlation. The lack of a detectable redshift correlation for $N_{HI}$ is in contrast with the clear redshift correlation for $N_{HX}$. Clearly, this does not provide support for the argument that redshift evolution in the GRB host properties is responsible for the redshift correlation for $N_{HX}$. (See Appendix A2 for further review of any $N_{HI}$ correlation with $N_{HX}$.

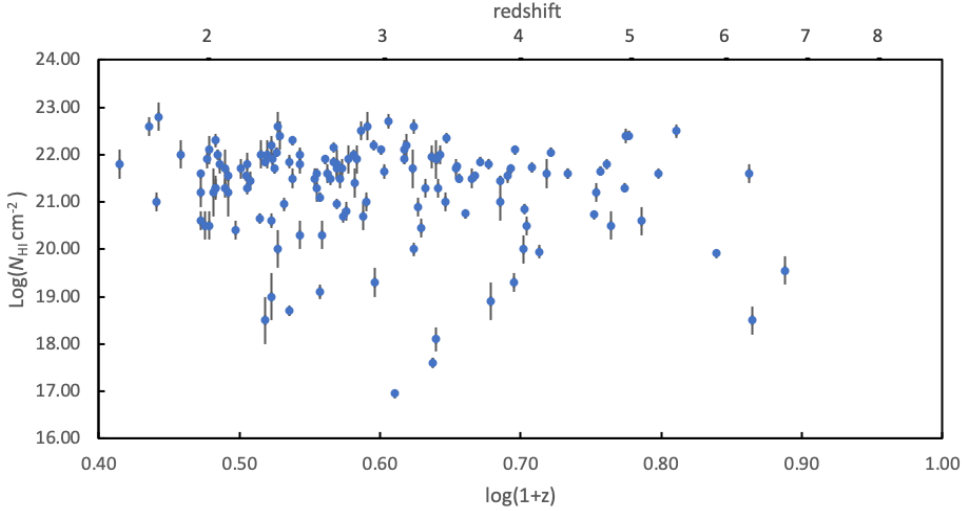

**Figure 2.** Distribution of $N_{HI}$ for GRB with redshift. No strong trends with redshift are visible.

### 3.3 USING $N_{HI,IC}$ AS PROXY FOR THE GRB INTRINSIC CONTRIBUTION TO $N_{HX}$

Part of the aim of this paper is to attempt to isolate an IGM contribution to $N_{HX}$. Here, we investigate the plausibility of assuming that the host intrinsic hydrogen column density is equal to the measured ionised corrected intrinsic neutral column ($N_{HI,IC}$) and examine the resulting residual column's dependence on redshift. To do this, we first make an ionisation correction (IC) to the hydrogen column density as measured by $N_{HI}$ using the approach described in Fumagalli, O'Meara and Prochaska, (2016) who observed that neutral fraction is a function of $N_{HI}$. The neutral fraction drops rapidly from ~ 0.7 at log($N_{HI}$ cm$^{-2}$) ~ 20 to ~0.02 at ~ log($N_{HI}$ cm$^{-2}$) ~ 18 with a 0.3 dex characteristic error. As the vast majority of GRB are in hosts with high column densities DLAs, only 11 out of 128 GRB sample required an IC.

Of the 128 GRB with data for both $N_{HX}$ and $N_{HI}$ in our sample, 96 have $N_{HX} > N_{HI,IC}$. In Figure 3, these are plotted, with the remaining 32 placed at the bottom of the figure for completeness. 32 of the GRB from this sub-sample (blue dots) are detections for both column densities. Where an estimate of $N_{HX}$ was available from the *Swift* repository but the confidence interval included zero, the object was plotted with a orange dot. Orange dots with arrows are the upper limits where *Swift* repository has a best fit of zero $N_{HX}$. The correlation statistics for the sub-sample of 32 GRB detections with $N_{HX} > N_{HI}$ are Pearson r=0.75 and Spearman ρ= 0.69 (for the full 128 sub-sample taking limits as detections one gets r=0.55, ρ=0.59.) The $N_{HX}$ – $N_{HI,IC}$ relation with redshift is much more significant than for a $N_{HX}$ alone. The reduced χ$^2$ =1.02 for a linear fit with the form $N_{HX}$ - $N_{HI,IC}$ α (1+z)$^{3.5±0.1}$ . It can be argued that this result supports the case for using $N_{HI,IC}$ as a proxy for the GRB intrinsic hydrogen column density, leaving the major remaining column density contribution being from the IGM. A caveat is that a large part of the sub-sample has been excluded as the X-ray or UV column density was not measured. The fraction excluded because $N_{HI,IC} > N_{HX}$ (32/128) is a cause for concern, as $N_{HX}$ is supposed to be the total column density. The large error bars on $N_{HX}$ may account for some of these. We can also see that the original $N_{HIGM}$ model is higher than the majority of the estimated intervening absorption. This may indicate that the IGM model is too simple e.g. it ignores LOS variation, or that the parameters used in the simple model need to be adjusted. However, the result could well be due to unrealistic assumptions of metallicity and ionisation for the GRB host galaxy. The next section examines the effects of assumptions about GRB host metallicity.

T. Dalton et al.

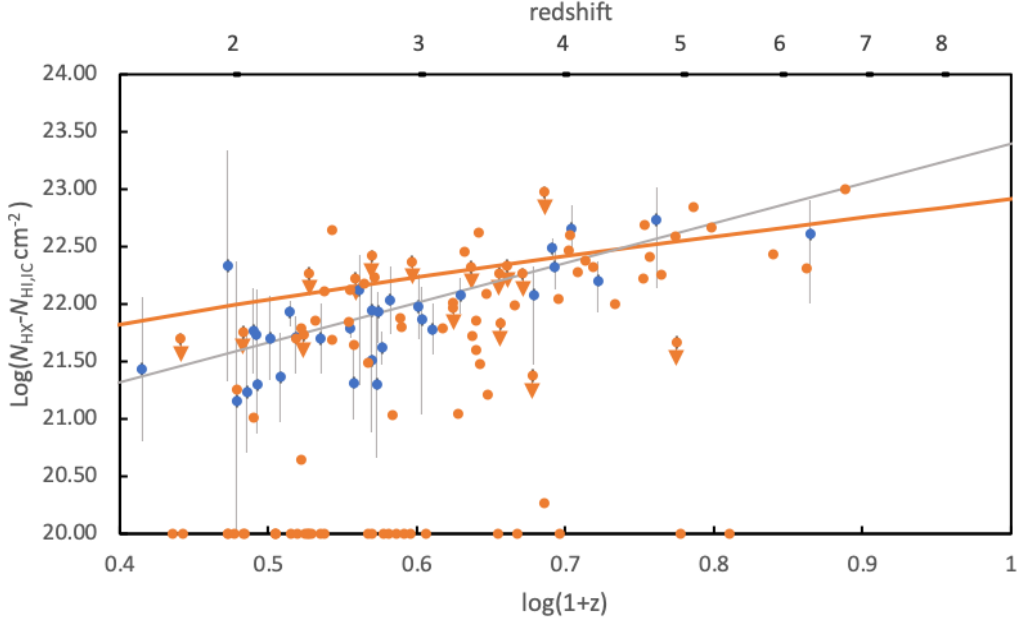

**Figure 3.** Distribution of total $N_{HX}$ minus the localised $N_{HI,IC}$ (which is being used as a proxy of the intrinsic GRB hydrogen column density) in the sub-sample of 128 GRB with data on both $N_{HX}$ and $N_{HI,IC}$. The blue dots are GRB detections for both $N_{HX}$ and $N_{HI,IC}$. The orange dots are objects for which the 90% confidence interval includes zero. Orange dots with arrows are the $N_{HX}$ upper limits where *Swift* repository has a best fit of zero. Where $N_{HX} - N_{HI,IC} < 0$ they are placed at 20.0 on the y-axis. The orange line is the integrated diffuse IGM $N_{HIGM}$, model described by equation (2). A power law fit to the $N_{HX} - N_{HI,IC}$ versus redshift trend scales as $(1+z)^{3.5+/-0.1}$ (grey line with reduced $\chi^2$ =1.02). Pearson and Spearman correlation coefficients are 0.75 and 0.69 respectively for the GRB detections, and 0.55 and 0.59 for the full sample where $N_{HX} > N_{HI,IC}$ *taking limits as detections*.

## 3.4  GRB HOST METALLICITY

GRB typically occur in sub-solar metallicity galaxy host environments (S13; Krühler et al., 2015; Cucchiara et al., 2015). Our sub-sample of 36 GRB with a range in redshift of $1.76 \leq z \leq 5.91$, and [X/H] from -2.18 to 0.25 is plotted in Figure 4. Where data from multiple sources were available, we took those with the smallest reported errors. Further, we only used data with detections and error bars, and excluded those with lower limits only. This resulted in omitting most of the 55 GRB from Cucchiara et al., (2015, hereafter C15). The Pearson and Spearman correlation coefficients are r = -0.24 and ρ = -0.27. However, both correlations fail the null hypothesis test, indicating there is no statistically significant correlation between GRB metallicity and redshift. The blue line is the best linear fit to the data is

$$[X/H] = (-1.01 \pm 0.04) - (0.09 \pm 0.01) z \qquad (4)$$

This possible mild metallicity evolution of GRB intrinsic gas with redshift is noted in some of the literature. For non-GRB absorption systems, stronger evolution is seen. For example, De Cia, Petitjean and Savaglio, (2018) reported evolution with a slope of ~0.32z. However, GRB metallicity does not appear to evolve as much, if at all, based on our sample. This is consistent with C15, for example.

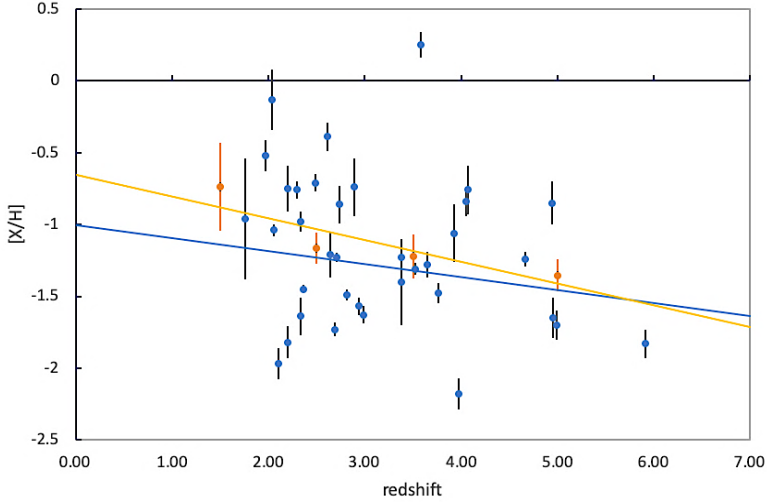

**Figure 4.** Distribution of the combined GRB absorption-based metallicities (blue dots) with redshift. The GRB absorption sample mean metallicity is -1.17 ± 0.09 (or 0.07 ± 0.05 $Z/Z_{sol}$). The orange dots are the weighted average metallicity over specific redshift bins of bins $\Delta z$ =1 (except for z=4 to 6 as there is only one GRB with z > 5) weighted by the total $N_{HI}$. The Pearson and Spearman correlation coefficients are r = -0.24 and ρ = -0.27, both correlations failing the null hypothesis tests, indicating there is no statistically significant correlation between GRB metallicity and redshift. The blue line is the $\chi^2$ linear fit to the blue dot data. The orange line is the best linear fit to the orange dots for the weighted average [X/H] and shows possible evolution (See Appendix A4 for more discussion).

It is well known that dust depletion affects the determination of metallicity in GRBs (e.g. Savaglio, 2006; De Cia et al., 2013). In Bolmer et al., (2019), 22 GRB are studied at z > 2 for features including dust depletion measurements and any relation to redshift. Based on this sample, they found that, on average, the dust corrected metallicity [M/H] = -1.09 ± 0.50 compared with -1.27 ± 0.37 for the uncorrected metallicity (0.08 $Z/Z_{sol}$ v 0.05 $Z/Z_{sol}$ respectively). This is an average correction of 0.2 dex which is considerably lower than found by De Cia et al., (2018) (0.4-0.5 dex) for non-GRB objects. In Figure 5, we plot the dust correction [M/H]-[X/H] versus redshift for the Bolmer et al., (2019) sample, to see if there is any obvious evolution with redshift. No detectable evolution is seen. The Pearson and Spearman coefficients for Figure 5 are -0.14 and 0.03 respectively and both fail the null hypothesis tests for a significant correlation.

Where actual dust corrections are not available, an argument can be made for using a standard dust correction to metallicity for XSPEC fitting for example, based on the Bolmer et al., (2019) mean value of 0.2 dex. This mean correction increases the average metallicity in our sample from 0.07 to 0.11 $Z/Z_{sol}$. While this is an important correction to [X/H], the impact on revised $N_{HX}$ is very small as the corrected metallicity is still <<solar [X/H]. Testing a sample of GRBs at redshift from 1 to 7, the change in log($N_{HX}$) after making a dust correction was 0.03 to 0.06 dex. Further, how any dust correction is estimated and used in the literature is not always clear. In conclusion, given that the impact of an average dust correction to $N_{HX}$ is very small, we do not consider that a standard dust correction to the metallicity adjustment is appropriate.

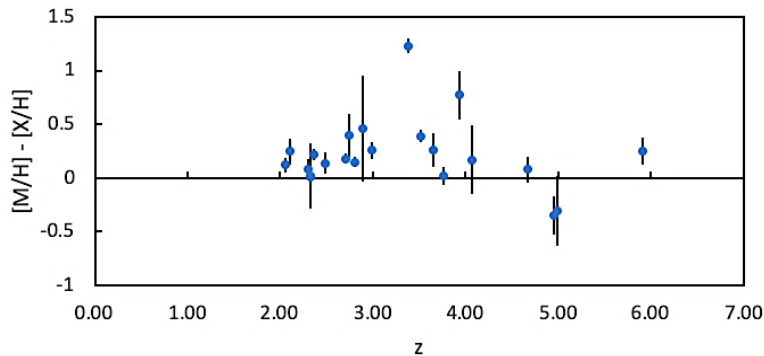

**Figure 5.** Plot of the dust correction to metallicities [M/H]-[X/H] by redshift for the *Bolmer* et al., *(2019)* GRB sample. The Pearson and Spearman coefficients are -0.14 and 0.03 respectively and both fail the null hypothesis tests for a significant correlation. There is no detectable evolution

.
In conclusion, from our GRB metallicity review, the GRB absorption sample mean metallicity is equal to -1.17 ± 0.09 or (0.07 ±0.05 Z/$Z_{sol}$). In further analysis therefore, we use the actual metallicity, dust corrected, for detections where available. While noting that some of the literature claims that there is possible mild redshift evolution in GRB host absorption, for the reasons outlined, we chose to use the average metallicity, without evolution or dust correction, of 0.07 Z/$Z_{sol}$ for the remaining GRB. This is certainly a more realistic value than simply assuming solar metallicity in revisiting the $N_{HX}$ for the full GRB sample in the next section.

### 3.5   IMPACT OF METALLICITY ASSUMPTIONS ON $N_{HX}$

We wish to examine the impact on GRB $N_{HX}$ fits in XSPEC of using actual dust corrected metallicities for GRB detections where available. For the remaining GRB, we examine a more realistic average host metallicity than solar, and importantly, look at the variation with redshift. To do this, we used an XSPEC model *tbabs\*tbvarabs\*po* for the X-ray data from GRB151027A (a very high S/N GRB), varying the modelled host redshift between 0 and 10 and testing for metallicities Z/$Z_{sol}$ = 0.07 (the mean from our sample in Section 3.4), and solar. A lower metallicity results in an increased fitted $N_{HX}$, with the increase varying with redshift (see Appendix A3 for more details). In order to to see whether this correction is consistent for different GRB X-ray spectra, we plotted the fractional increase in fitted $N_{HX}$ with redshift for a test sample of three high S/N GRB spectra with differing reported redshifts and $N_{HX}$. Again, we varied the redshifts between 0 and 10 and used metallicity = 0.07 Z/$Z_{sol}$ compared with the value assuming solar metallicity. The fractional increase in $N_{HX}$ with redshift is very similar for the three GRB). A power law fit to the increase for GRB151027A is (orange line in Appendix Figure A3.2)

$$\Delta\log(N_{HX}\text{ cm}^{-2}) = (0.59 \pm 0.04)\log(1+z) + 0.18 \pm 0.02 \quad (5)$$

A more accurate power law could be obtained from a combined fit for the three GRB. However, this fit is deemed sufficient for the purposes of analysing the impact of a more realistic general metallicity assumption when calculating $N_{HX}$.

### 3.6   GRB $N_{HX}$ REVISED FOR REALISTIC HOST METALLICITY

Using actual metallicities, dust corrected where available, and the above power law relation for the remaining GRB, we use the new $N_{HX}$ for our full GRB sample and replot the relation with redshift in Figure 6.

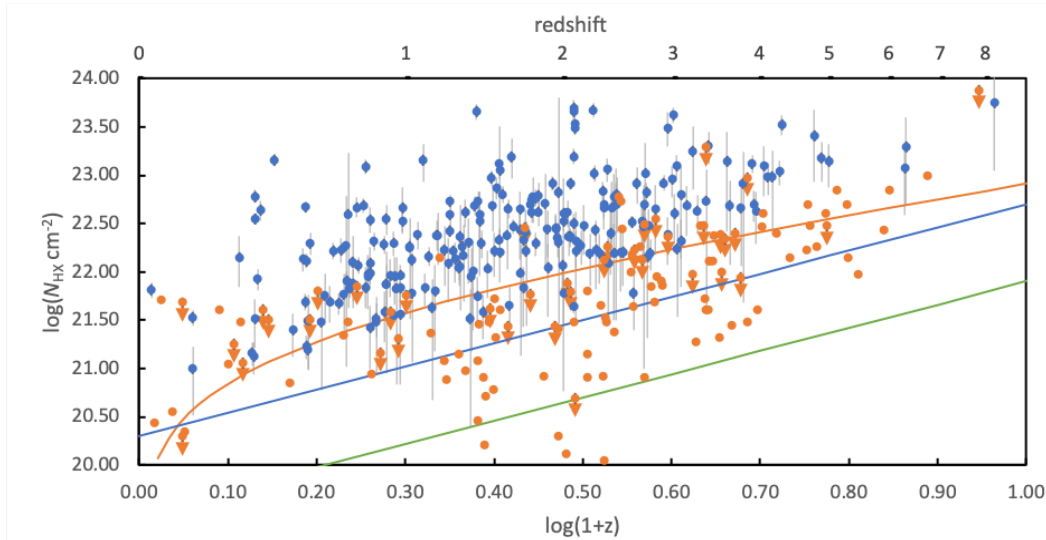

**Figure 6.** GRB $N_{HX}$ revised using actual metallicities, dust corrected where available and a mean metallicity of 0.07 Z/$Z_{sol}$ for the remaining GRB . Blue dots are GRB detections. The orange dots are objects for which the 90% confidence interval includes zero. Orange dots with arrows are the upper limits where the *Swift* repository has a best fit of zero. The orange line is the simple model IGM line from equation (2), $N_{HIGM}$. The blue line is the GRB lower envelope based on the requirement for 90% of detections, including error bars to be above the envelope, with log ($N_{HX}$ (z=0) cm$^{-2}$) = 20.3. The green line is the envelope with the requirement that 99% of all GRB, ignoring error bars, are above the envelope, which has log($N_{HX}$ (z=0) cm$^{-2}$) = 19.5 following the rule of thumb in *Campana* et al*., (2015)*. Both these envelopes are plotted with assumed slopes of $(1+z)^{2.4}$ (Campana et al., 2014). The Pearson and Spearman correlation coefficients are 0.59 and 0.61 respectively for GRB detections and 0.62 and 0.62 for the full sample treating the limits as detections.

The Pearson and Spearman correlation coefficients are 0.59 and 0.61 respectively for GRB detections in Figure 6, which are stronger than prior to the correction for a low metallicity. Blue dots are GRB detections. Orange dots with arrows are upper limits where the fitted $N_{HX}$ are 0, and orange dots are where the 90% confidence interval includes zero. The orange line in Figure 6 is the simple model IGM line from equation (2). The blue line is an estimate of the GRB lower envelope based on a requirement of having 90% of detections including their error bars to be above the envelope.

$$\log (N_{HX}\ cm^{-2}) = 20.3 + 2.4\log(1+z) \quad (6)$$

The green line is an estimate of the lower envelope based on the requirement that 99% of all GRB measurements, ignoring error bars and treating any upper limits as detections, are above the envelope, using the rule of thumb in Campana et al., (2015) (note that they put $N_{HX}$ at the top of the 90% confidence interval where the 90% confidence interval of a fit includes zero, whereas we use the Swift best estimate $N_{HX}$ which is lower, except for those with best estimates equal to zero).

$$\log (N_{HX}\ cm^{-2}) = 19.5 + 2.4\log(1+z) \quad (7)$$

The envelope fits may give an indication of the maximum $N_{HIGM}$ potential contribution to $N_{HX}$. Both the envelopes have been assumed to scale with redshift as $(1+z)^{2.4}$ (Campana et al., 2014), and we note that this may only be realistic for a cold absorber and not for a warm absorber (S13). The GRB LOS goes through a wide range of environments with different temperatures and densities. This will change the effective absorption cross-section at different frequencies. However, we will retain the cold absorber approximation for the current analysis. Using a $\chi^2$ fit, the revised $N_{HX}$ for the detections scale as $N_{HX} \alpha (1+z)^{1.94\pm0.04}$. However, a large reduced $\chi^2$ indicates that the relationship is not a simple power law or that the data has a large additional source of scatter. We explore this further in Section 3.7. Of 226 GRB detections, only 11 are now below the $N_{HIGM}$ curve, not taking error bars into account.

### 3.7 Revised GRB $N_{HX}$ - $N_{HI,IC}$

As before, to isolate the IGM contribution to $N_{HX}$, we subtract from the revised $N_{HX}$ the GRB $N_{HI}$ adjusted for an ionisation correction, as a proxy for intrinsic hydrogen column density, as we did in Section 3.3, and plot the result against redshift in Figure 7.

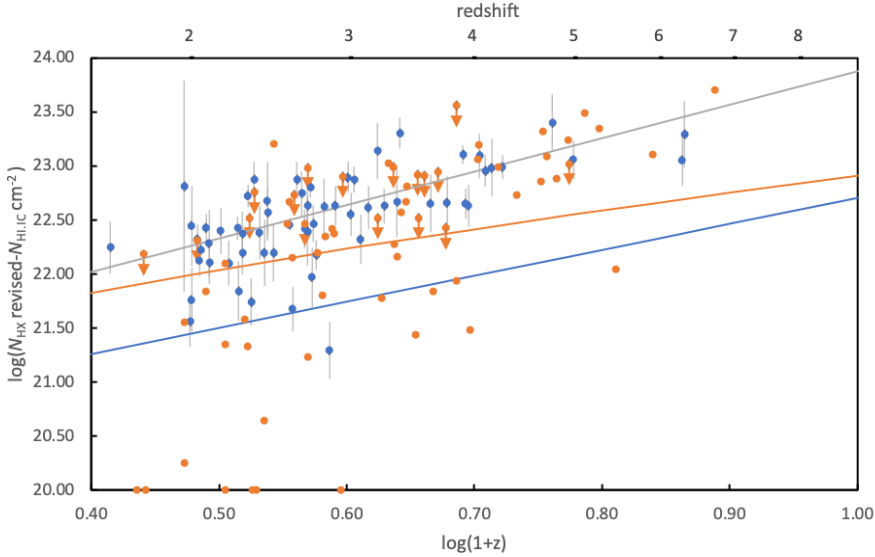

**Figure 7.** Distribution of GRB revised $N_{HX}$ - $N_{HI,IC}$ with redshift. The blue dots are GRB detections and the orange with arrows are upper limits where the $N_{HX}$ best fit is zero, or orange dots where the 90% confidence interval of the x-ray fit includes zero. Where $N_{HX} < N_{HI,IC}$ the GRB are placed at the bottom of the figure for completeness. The blue line is the GRB lower envelope based on the requirement for 90% of detections, including error bars to be above the envelope, and has $\log(N_{HX}(z=0)\ cm^{-2})=20.3$. The orange line is the original simple model $N_{HIGM}$ for $n_0 = 1.7 \times 10^{-7} cm^{-3}$. The $N_{HX}$ - $N_{HI,IC}$ for GRB detections best fit has a power law slope of $(1+z)^{3.1\pm-0.3}$ (grey line, reduced $\chi^2 =2.6$). The Pearson and Spearman correlation coefficients are 0.65 and 0.67 respectively for the GRB detections, and 0.53 and 0.62 for the full sample with limits treated as detections and where $N_{HX} < N_{HI,IC}$.

Of the 128 GRB with data for both $N_{HX}$ and $N_{HI}$ in our sample, 122 now have $N_{HX} > N_{HI,IC}$. The $N_{HX}$ - $N_{HI,IC}$ for GRB detections now has a best fit power law slope of $(1+z)^{3.1+/-0.3}$ (grey line – with reduced $\chi^2$ =2.6). The Pearson and Spearman correlation coefficients are 0.65 and 0.67 respectively for the GRB detections, and 0.53 and 0.62 for the full sample with best fits being treated as detections, where upper limits are treated as detections where the best fit equals zero, and where $N_{HX} > N_{HI,IC}$.

This final figure is our best representation of the use of GRB X-ray spectral fits to potentially constrain the IGM hydrogen column density. The blue line is the GRB lower envelope based on the requirement for 90% of detections (equation (6)), including error bars, to be above the envelope. Using equation (2), the orange line is $N_{HIGM}$ for a mean hydrogen number density $n_0 = 1.7 \times 10^{-7}$ cm$^{-3}$.

In Figure 3, 32 GRB had $N_{HX} < N_{HI,IC}$ compared to only 6 based on the revised $N_{HX}$ for our updated GRB host metallicity in Figure 7. Therefore, the more realistic GRB metallicity generates a more plausible $N_{HX}$, if it is assumed to represent the total hydrogen column density, which hence must be greater than the intrinsic column density. Given this requirement, we examined the 6 GRB where the fitted $N_{HX}$ was less than $N_{HI,IC}$. For these objects, Table 2 lists: 1) GRB name; 2) log($N_{HI}$ cm$^{-2}$); 3) Measured metallicity from literature if available; 4) revised log($N_{HX}$ cm$^{-2}$); 5) Whether the revised $N_{HX} > N_{HI,IC}$); 6) log($N_{HIGM}$ cm$^{-2}$) at the GRB redshift; and 7) whether the revised $N_{HX}$ - $N_{HI,IC}$ is greater than $N_{HIGM}$. We also include GRB180624A in the table, which had a revised $N_{HX}$ - $N_{HI,IC}$ substantially below $N_{HIGM}$.

We refitted each GRB in XSPEC using *tbabs*tbvarabs*po* using the actual reported metallicity, or 0.07 Z/Z$_{sol}$ otherwise. As can be seen from Table 2, all GRB now have $N_{HX} > N_{HI,IC}$. Further, all show $N_{HX}$ less $N_{HI,IC}$ (GRB160227A within error bars) as proxy for the host intrinsic column density, being greater than $N_{HIGM}$. The refitting using the actual redshift and metallicity (or 0.07 otherwise) gives a higher corrected $N_{HX}$ as the power law correction approximation marginally understates the actual relation between metallicity correction and redshift for redshift between 0.3 < log(1+z) < 0.8 (See Appendix A3). Of the 67 GRB detections, 5 lie below the $N_{HIGM}$ curve in Figure 7. None lie below the $N_{HIGM}$ curve after refitting for more realistic or actual metallicity.

In conclusion, by using actual metallicities, dust corrected where available, and a more realistic average GRB metallicity than the standard solar assumption for the remainder, we have shown that the revised larger $N_{HX}$ is greater than an ionisation corrected $N_{HI}$ for our entire sample of 128 GRB, where measurements of both are available, together with a spectroscopic redshift. Further, the lower envelope of $N_{HX}$ - $N_{HI,IC}$ is potentially a useful constraint on the IGM contribution to $N_{HX}$. Finally, the metallicity revised $N_{HX}$ - $N_{HI,IC}$ for detections are mostly above the simple model $N_{HIGM}$ curve further suggesting that this is a useful constraint on the IGM hydrogen column density.

**Table 2.** Summary analysis for refitting of 6 GRB where the revised $N_{HX}$ <, $N_{HI,IC}$ and one where the revised $N_{HX}$ - $N_{HI,IC}$ is substantially below $N_{HIGM}$. 1) GRB name; 2) log($N_{HI}$ cm$^{-2}$); 3) Measured metallicity from literature if available; 4) revised log($N_{HX}$ cm$^{-2}$); 5) Whether the revised $N_{HX} > N_{HI,IC}$); 6) log($N_{HIGM}$ cm$^{-2}$) at the GRB redshift; and 7) whether the revised $N_{HX}$ - $N_{HI,IC}$ is greater than $N_{HIGM}$.

| GRB | log($N_{HI}$ cm$^{-2}$) | Measured Z/Z$_{sol}$ | Log($N_{HX}$ cm$^{-2}$) | New $N_{HX} > N_{HI,IC}$? | log($N_{HIGM}$ cm$^{-2}$) at GRB z | $N_{HX}$ - $N_{HI,IC}$>IGM? |
|---|---|---|---|---|---|---|
| 050922C | 21.55+/-0.10 | 0.15[1] | 22.04* | Y | 22.04 | Y |
| 120119A | 22.60+/-0.20 | 0.11[2] | 22.45+/-1.74 | Y | 21.91 | Y |
| 120815A | 22.05+/-0.10 | 0.04[3] | 22.16* | Y | 22.09 | Y |
| 121027A | 22.80+/-0.30 | ** | 22.86+/-1.20 | Y | 21.92 | Y |
| 160227A | 22.40+/-0.20 | ** | 22.08+/-1.15 | Y | 22.10 | Within error bars |
| 181020A | 22.20+/-0.10 | 0.27[3] | 22.38+/-1.25 | Y | 22.22 | Y |
| 180624A | 22.5+/-0.20 | ** | 22.70+/-3.40 | Y | 22.21 | Y |

Notes: Z/Z$_{sol}$ references: [1] (Arabsalmani et al., 2018), [2] (Heintz et al., 2019), [3] (Bolmer et al., 2019), * log($N_{HX}$) lower error bar not with 90% confidence, ** metallicity unknown so 0.07 Z/Z$_{sol}$ used in XSPEC fitting

**CONCLUSION**

We compiled a large sample of all *Swift* X-ray Telescope observed GRB with spectroscopic redshifts up to 31 July 2019, (with a photometric redshift only for GRB090429B). Of this sample of 352 GRB with fitted X-ray equivalent hydrogen column densities, 128 have also have intrinsic neutral hydrogen column density measurements. We have also compiled a sample of absorption-based metallicity data. The main aims of this paper are to generate improved $N_{HX}$ for our sample by using more realistic host metallicity and, by approximating the host intrinsic hydrogen column density as equal to the measured $N_{HI}$, with an ionisation correction applied, to isolate the more accurate IGM column density contribution.

We analysed the impacts on $N_{HX}$ of using metallicities that more realistically reflect the LOS absorption through the host galaxy than the standard use of the solar abundance. We discussed the possibility of using an average dust correction where actual measurements were not available but it had an insignificant effect on $N_{HX}$.

Our main findings and conclusions are:
1. While some of the literature notes that GRB metallicity shows a mild evolution with redshift, the Pearson and Spearman correlation coefficients for our sample are -0.24 and -0.27 respectively, and both correlations fail the null hypothesis test, indicating there is no detected trend. Further, the large reduced $\chi^2$ of the fit means either that a linear model is not a good description of any potential relation or that there is a large additional source of scatter. Hence we do not find a statistically significant relation between GRB metallicity and redshift.
2. The GRB absorption sample mean metallicity is [X/H] = -1.17 ± 0.09 (or 0.07 ± 0.05 $Z/Z_{sol}$). This is substantially lower than the assumption of solar metallicity used as standard for many fitted $N_{HX}$.
3. We find that using a lower GRB host metallicity results in increasing the fitted $N_{HX}$ with the correction scaling with redshift. In order to determine this relation at mean metallicity 0.07 $Z/Z_{sol}$, we plotted the fractional increase in $N_{HX}$ with redshift for some trial fits. We find that the fractional increase in $N_{HX}$ with redshift is very similar for a range of GRB fits. The power law relation for GRB151027A, used as a standard GRB for metallicity 0.07 $Z/Z_{sol}$ is $\Delta\log(N_{HX}$ cm$^{-2}$) = (0.59 ± 0.04) log(1+z) + (0.18 ± 0.02).

   A more accurate power law could be obtained from a combined fit of a large sample of GRB. However, this is sufficient for our purposes of analysing the impacts of a more realistic general metallicity assumption for calculating $N_{HX}$.
4. Using actual metallicities, dust corrected where available, and, for the remaining GRB, our power law relationship for the mean GRB host metallicity of 0.07 $Z/Z_{sol}$, we revised the $N_{HX}$ for our full GRB sample and replotted the relation with redshift. To more accurately isolate the IGM contribution to the total hydrogen column density, we subtracted from the revised $N_{HX}$ the GRB $N_{HI}$ after ionisation correction, as a proxy for the intrinsic hydrogen X-ray column density, and plotted the result against redshift. Of the 128 GRB with data for both $N_{HX}$ and $N_{HI}$ in our sample, only 6 have $N_{HI,IC}$ greater than the revised $N_{HX}$, compared to 32 when solar metallicity is assumed. Therefore, using more realistic GRB metallicities generates an improved $N_{HX}$, if it is interpreted as representing the total hydrogen column density, which must be greater than the local neutral column density. The estimated $N_{HX}$ - $N_{HI,IC}$ for GRB detections now has a redshift dependence of $(1+z)^{3.1+/-0.3}$ for the GRB detections, compared with power laws of 3.5+/-0.1 for $N_{HX}$ fitted assuming solar abundance. The Pearson and Spearman correlation coefficients are 0.65 and 0.67 respectively for the GRB detections, and 0.53 and 0.62 respectively for the full sample where $N_{HX}$ > $N_{HI,IC}$.
5. The lower envelope of the revised $N_{HX}$ > $N_{HI,IC}$ plotted against redshift has $N_{HX}$ (z=0) = 20.3 cm$^{-2}$ for our GRB sample of revised $N_{HX}$ - $N_{HI,IC}$. This is taken to be representative of the maximum IGM hydrogen column density, based on the requirement for 90% of detections, including error bars, to be above the envelope. Using this approach, we estimate the IGM $n_0$ to be equal to
   1.7x10$^{-7}$cm$^{-3}$ for the $N_{HIGM}$ curve which is consistent with that used by Behar et al., (2011) and S13.

X-ray spectroscopy at higher resolution and at higher signal-to-noise than is currently available would be required to detect absorption edges from individual ions in GRB. Such observations in the future, will provide valuable data on the distribution of the material along the line of sight, including its temperature, composition, density and ionisation state. The value found for the column density is almost always determined assuming a 100% neutral absorbing gas. This neutral assumption would cause the $N_{HX}$ to be underestimated if incorrect. Therefore, we can further improve the $N_{HX}$ and GRBs as probes of the IGM when higher resolution X-ray spectroscopy becomes available. We plan to examine the properties of the IGM such as metallicity, temperature and density in a subsequent paper to develop a better IGM model and compare it with the results of this paper.

T. Dalton et al.

## ACKNOWLEDGEMENTS

We thank M. Fumagalli for useful comments, K. Page for information on the *Swift* repository, and R. Mahmaud and K. Arnaud for advice on XSPEC. We also thank the referee for providing valuable and constructive feedback. S.L. Morris also acknowledges support from STFC (ST/P000541/1).

T. Dalton et al.

## Appendix

*A.1 S/N bias review*

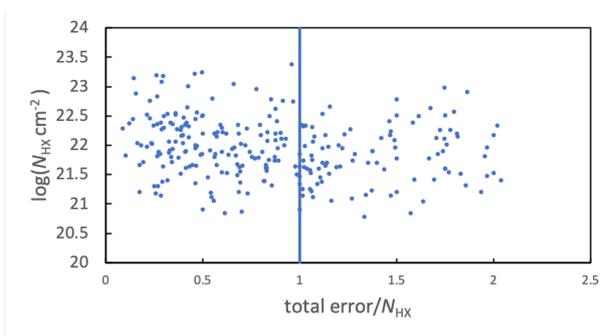 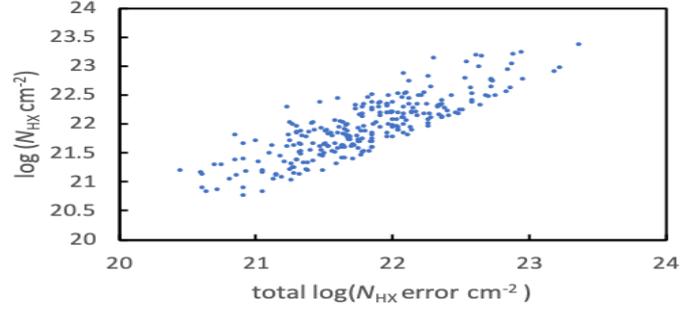

**Figure A1.** The left panel shows the X-ray equivalent hydrogen column density against the ratio of the total error/$N_{HX}$. The blue vertical line is where total error/$N_{HX}$ is 1. The scatter appears random. The right panel shows the distribution of $N_{HX}$ against the total error for the detection and shows a strong correlation. A S/N limited sample based on total error would introduce a bias to low $N_{HX}$ GRB values.

As we are mainly using *Swift* detected GRB where a redshift is available, we wished to examine if a S/N limited sample would cause bias. As a proxy for S/N, we plotted the log($N_{HX}$) versus both log of total error in $N_{HX}$ and total error/$N_{HX}$ for all detections in Figure A1. The left panel plots the X-ray equivalent hydrogen column density against the ratio of the total error/$N_{HX}$ where the total error is the 90% confidence range of the $N_{HX}$ fit. The scatter appears random so any cut-off
by total error/$N_{HX}$ should not result in a bias in $N_{HX}$. The right panel plots distribution of $N_{HX}$ against the total error for the detection. There a clear strong correlation so any cut-off for a S/N limited sample based on total error could produce a bias towards low $N_{HX}$ GRB values. As a test of the impact of using a flux limited sample, we restricted a sub-sample to total error/$N_{HX}$ < 1 resulting in 163 GRB.

This sub-sample had essentially the same properties as the GRB detection sample using the Pearson and Spearman correlation coefficients and the $N_{HX}$ versus redshift trend. Based on these results, we chose to use our full samples and not limit by a minimum flux.

*A.2 $N_{HX}$ and $N_{HI}$ correlation review*

In Figure A2, we plot $N_{HX}$ and $N_{HI,IC}$ for the full sub-sample of 128 GRB with both $N_{HX}$ and $N_{HI}$ data.
No strong correlation is detected using a null hypothesis test, with both Pearson and Spearman coefficients being 0.10. It can be argued that this result strengthens the case that it is the IGM that is causing the $N_{HX}$ redshift relation and is not intrinsic to the GRB host.

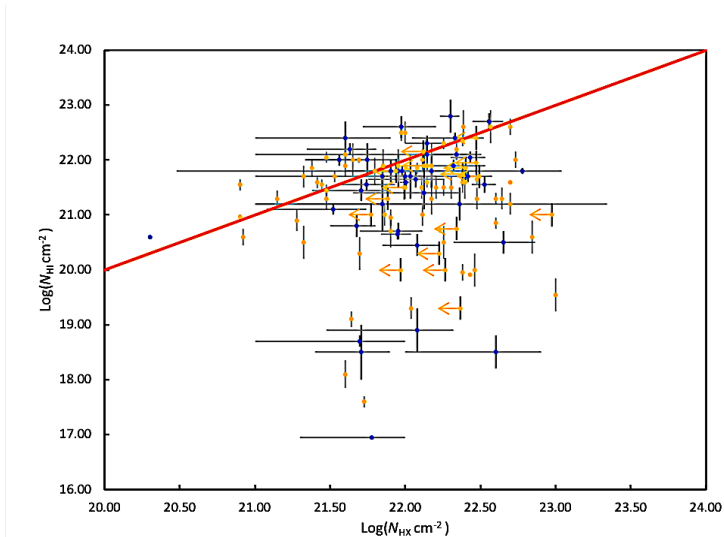

**Figure A2.** Plot of the log of the column densities for 128 GRB with both $N_{HX}$ and $N_{HI}$ data from Tanvir et al., (2019) with ionisation corrections . Blue dots are GRB for which the $N_{HX}$ are detections. The orange dots are GRB best fits per Swift but where the 90% confidence includes zero, and orange dots with arrows where the best fit $N_{HX}$ = zero. The line shows where $N_{H(,IC}$ is equal to $N_{HX}$. There is no correlation.

T. Dalton et al.

## A.3 Power law for Metallicity redshift scaled adjustment to $N_{HX}$

To examine the impact on GRB $N_{HX}$ fits in XSPEC using a more realistic mean GRB host metallicity than solar, and the variation with redshift, we used an XSPEC model *tbabs\*tbvarabs\*po* for GRB151027A (a high S/N GRB. We varied the redshifts between 0 and 10 and used metallicity = 0.07 $Z/Z_{sol}$ in fitting $N_{HX}$. Figure A3.1 shows a clear increasing metallicity adjusted log($N_{HX}$) with redshift (blue line for Z = 0.07). In order to determine this relationship, and see whether it is consistent for different GRB X-ray spectra, we plotted in Figure A3.2 the fractional increase in $N_{HX}$ with redshift for a test sample of three high S/N GRB spectra with differing reported redshifts and $N_{HX}$. The fractional increase in $N_{HX}$ with redshift is very similar for the three GRB. The power law relation from a best fit least squares for GRB151027A is log($N_{HX}$) = (0.59 ± 0.04)log(1+z) + (0.18 ± 0.02).

We note that in Figure A3.2, the power law curve is higher at log(1+z)<0.2 and log(1+z) >0.8. As a result, using this relation to adjust $N_{HX}$ will result in marginally higher values than actual at low and high redshift but is a reasonable approximation for our purposes.

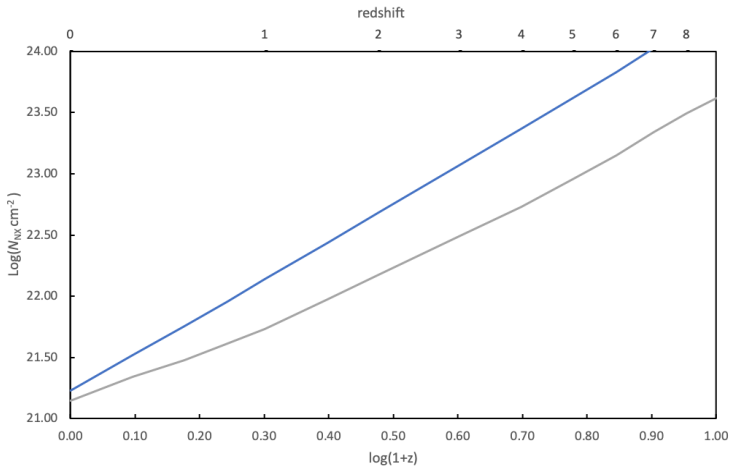

**Figure A3.1** Impact on fitted $N_{HX}$ for GRB151027A varying the host redshift between 0 and 10 and using metallicities of 0.07 $Z/Z_{sol}$ (mean from our sample in Section 3.4), and solar. The blue line is with Z=0.07 and the grey Z=1. A lower metallicity results in increasing the fitted $N_{HX}$ with the increase varying with redshift.

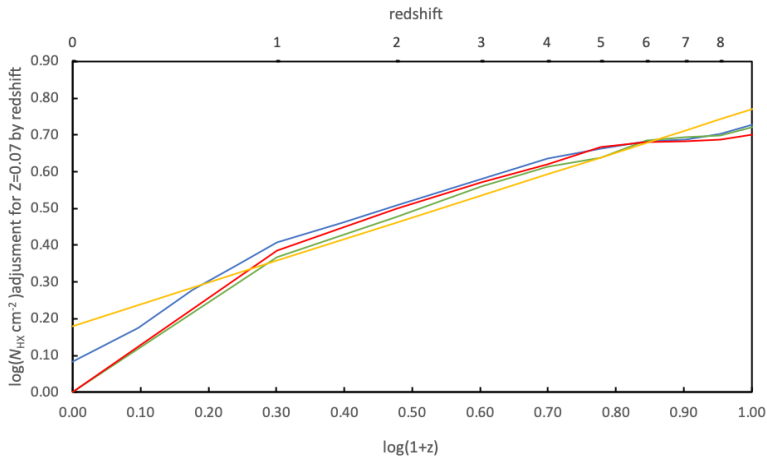

**Figure A.3.2** Comparison of the fractional increase in $N_{HX}$ with redshift for three high S/N GRB, blue is GRB151027A, red is GRB150403A and green is GRB120909A. The power law relation yellow line is from a least squares best fit for GRB151027A, log($N_{HX}$) = (0.59 ± 0.04)log(1+z) +(0.18 ±0.02).

.

### A.4 $N_{HI}$ weighted average metallicity of GRB hosts (See Figure 4)

The sample of metallicity measurements can be used to investigate the cosmic metallicity at different redshifts. A method to do this is to weight the average metallicity over a specific redshift bin with the total neutral hydrogen column density in the same redshift interval (C15). We used redshifts bins of $\Delta z = 1$ except for redshift 4 to 6 as there was only one GRB datapoint for metallicity greater than redshift 5. This weighted sample gives a marginally stronger evolution for metallicity than for the individual GRB.

$$[X/H]_{N_{HI}\ weighted} = -0.65\pm0.07 - (0.15\pm0.06)z \quad (A1)$$

Despite this possible mild evolution, in the high redshift range z >4, all the GRB are metal enriched from 0.02-0.17 $Z/Z_{sol}$, suggesting that substantial amounts of metals were already present in galaxies the early universe. The lack of clear redshift evolution is in contrast with quasars (QSO) which show strong evolution (Rafelski et al., 2014). A possible explanation for the lack of evolution is that GRBs may be located in different environments to quasars (Fynbo *et al.*, 2008; C15). However, while this would affect the emission line metallicity, it should impact less on the absorption metallicity which is tracing the average galaxy LOS (Arabsalmani et al., 2018). We need far more GRB at high redshift to increase the statistical significance of metallicity evolution. For this paper, we will not use the $N_{HI}$ weighted average values as we wish to establish a metallicity to be applied to each GRB for the $N_{HX}$ fitting.

# Table 1 Full On-line version

**Reference detail**

| Ref | Full reference |
|---|---|
| Swift | www.swift.ac.uk/xrt_spectra |
| A16 | Arcodia, R., Campana, S. and Salvaterra, R. (2016)A&A, *590*, pp. 1–8. |
| S15 | Salvaterra, R., 2015.  Journal of High Energy Astrophysics, *7*, pp.35-43. |
| Z18 | Zafar, T. *et al.* (2018) MNRAS, 479(2), p. 1542. |
| LJ15 | Littlejohns, O.M., Butler, N.R., Cucchiara, A., et al., 2015, MNRAS, *449*(3), pp.2919-2936. |
| Tanvir19 | Tanvir, N.R., Fynbo, J.P.U., de Ugarte Postigo, et al., 2019, MNRAS, *483*(4), pp.5380-5408. |
| A18 | Arabsalmani, M., Møller, P., Perley, D.A., et al., 2018, MNRAS, *473*(3), pp.3312-3324. |
| C15 | Cucchiara, A., Fumagalli, M., Rafelski, M., et al., 2015, AJ, *804*(1), p.51. |
| V17* | Vergani, S.D., Palmerio, J., Salvaterra, R., et al, 2017,A&A, *599*, p.A120. |
| H19 | Heintz, K.E., Bolmer, J., Ledoux, C., Noterdaeme, et al., 2019. A&A, *629*, p.A131. |
| B19 | Bolmer, J., Ledoux, C., Wiseman, P., et al.,2019, A&A, *623*, p.A43. |
| IC | Ionisation Correction from Fumagalli, M., O'Meara, J. M. and Xavier Prochaska, J. (2016)', MNRAS, 455(4), pp. 4100–4121. |

| GRB | z | log($N_{HX}$cm$^{-2}$) | + | - | ref | log($N_{HI}$cm$^{-2}$) | +/- | IC | [X/H] | + | - | Ref Tanvir19 |
|---|---|---|---|---|---|---|---|---|---|---|---|---|
| 000301C | 2.03 | | | | | 21.20 | 0.50 | | | | | |
| 000926 | 2.04 | | | | | 21.30 | 0.25 | | -0.13 | 0.21 | | A18 |
| 011211 | 2.14 | | | | | 20.40 | 0.20 | | -1.22 | min | | C15 |
| 020124 | 3.20 | | | | | 21.70 | 0.40 | | | | | |
| 021004 | 2.33 | | | | | 19.78 | 0.50 | IC | | | | |
| 030226 | 1.99 | | | | | 20.50 | 0.30 | | -1.28 | min | | C15 |
| 030323 | 3.37 | | | | | 21.90 | 0.07 | | -1.32 | min | | C15 |
| 030429 | 2.65 | | | | | 21.60 | 0.20 | | -1.13 | min | | C15 |
| 050126 | 1.29 | 19.90 | 1.92 | 19.90 | Swift | | | | | | | |
| 050223 | 0.59 | 21.80 | 0.00 | 21.80 | Swift | | | | | | | |
| 50315 | 1.95 | 21.99 | 0.12 | 0.15 | Swift | | | | | | | |
| 050318 | 1.44 | 20.90 | 0.43 | 20.90 | Swift | | | | | | | |
| 050319 | 3.24 | 21.28 | 0.54 | 21.28 | Swift | 20.90 | 0.20 | | -0.77 | min | | C15 |
| 050401 | 2.90 | 22.39 | 0.06 | 0.07 | A16 | 22.60 | 0.30 | | -1.07 | min | | C15 |
| 050408 | 1.24 | 22.20 | 0.10 | 0.12 | Swift | | | | | | | |
| 050416A | 0.65 | 21.91 | 0.07 | 0.08 | A16 | | | | | | | |
| 050505 | 4.27 | 22.43 | 0.10 | 0.13 | Swift | 22.05 | 0.10 | | -1.20 | min | | C15 |
| 050509B | 0.23 | 21.60 | 0.51 | 21.60 | Swift | | | | | | | |
| 050525A | 0.61 | 21.18 | 0.27 | 0.70 | Swift | | | | | | | |
| 050603 | 2.82 | 22.54 | 0.00 | 22.54 | Swift | | | | | | | |
| 050724 | 0.26 | 21.04 | 0.46 | 21.04 | Swift | | | | | | | |
| 050730 | 3.97 | 21.60 | 0.30 | 21.60 | Swift | 22.10 | 0.10 | | -1.96 | 0.11 | 0.11 | C15 |
| 050802 | 1.71 | 21.40 | 0.21 | 0.40 | Swift | | | | | | | |
| 050803 | 4.30 | 22.91 | 0.09 | 0.10 | Swift | | | | | | | |

| | | | | | | | | | | | |
|---|---|---|---|---|---|---|---|---|---|---|---|
| 050814 | 5.30 | 22.15 | 0.42 | 22.15 | Swift | | | | | | |
| 050820A | 2.61 | 21.52 | 0.22 | 0.44 | Swift | 21.10 | 0.10 | | -0.39 | 0.10 | 0.10 | A18 |
| 050823 | 1.43 | 21.89 | 0.12 | 0.15 | Swift | | | | | | |
| 050824 | 0.83 | 20.94 | 0.43 | 20.94 | Swift | | | | | | |
| 50826 | 0.30 | 21.90 | 0.21 | 0.30 | Swift | | | | | | |
| 050904 | 6.29 | 22.38 | 0.23 | 0.48 | Swift | 21.60 | 0.20 | | | | |
| 050908 | 3.34 | 21.72 | 0.74 | 21.72 | Swift | 19.48 | 0.30 | IC | | | |
| 050922C | 2.20 | 20.90 | 0.74 | 20.90 | Swift | 21.55 | 0.10 | | -1.82 | 0.11 | 0.11 | A18 |
| 051001 | 2.43 | 22.28 | 0.23 | 0.38 | Swift | | | | | | |
| 051016B | 0.94 | 21.95 | 0.11 | 0.12 | Swift | | | | | | |
| 051022 | 0.80 | 22.76 | 0.05 | 0.05 | Swift | | | | | | |
| 051109A | 2.35 | 20.04 | 1.68 | 20.04 | Swift | | | | | | |
| 051111 | 1.55 | 21.78 | 0.22 | 0.30 | Swift | | | | | | |
| 051221A | 0.55 | 21.20 | 0.26 | 0.51 | Swift | | | | | | |
| 060115 | 3.53 | 21.99 | 0.00 | 21.99 | Swift | 21.50 | 0.10 | | -1.53 | min | | C15 |
| 060124 | 2.30 | 21.71 | 0.19 | 0.31 | Swift | 19.70 | 0.60 | IC | | | |
| 060202 | 0.79 | 22.36 | 0.06 | 0.06 | Swift | | | | | | |
| 060204B | 2.34 | 22.20 | 0.12 | 0.12 | Swift | | | | | | |
| 060206 | 4.05 | 22.60 | 0.40 | 22.60 | Swift | 20.85 | 0.10 | | -0.84 | 0.10 | 0.10 | A18 |
| 060210 | 3.91 | 22.53 | 0.07 | 0.08 | A16 | 21.55 | 0.15 | | -0.83 | min | | C15 |
| 060218 | 0.03 | 21.62 | 0.06 | 0.07 | Swift | | | | | | |
| 060223A | 4.41 | 22.15 | 0.91 | 22.15 | Swift | 21.60 | 0.10 | | -1.80 | min | | C15 |
| 060306 | 1.55 | 22.63 | 0.07 | 0.08 | Swift | | | | 0.43 | 0.18 | 0.42 | V17 |
| 060418 | 1.49 | 21.62 | 0.00 | 21.62 | Swift | | | | | | |
| 060502A | 1.52 | 21.80 | 0.18 | 0.27 | Swift | | | | | | |
| 060510B | 4.94 | 22.60 | 0.40 | 22.60 | Swift | 21.30 | 0.10 | | -0.84 | min | | C15 |
| 060512 | 2.10 | 20.69 | 0.00 | 20.69 | Swift | | | | | | |
| 060522 | 5.11 | 22.85 | 0.36 | 22.85 | Swift | 20.60 | 0.30 | | | | |
| 060526 | 3.21 | 21.97 | 0.00 | 21.97 | Swift | 20.00 | 0.15 | | | | |
| 060605 | 3.77 | 22.08 | 0.24 | 0.60 | Swift | 19.69 | 0.50 | IC | | | |
| 060607A | 3.08 | 21.78 | 0.22 | 0.48 | Swift | 18.91 | 0.30 | IC | | | |
| 060614 | 0.13 | 20.34 | 0.31 | 20.34 | A16 | | | | | | |
| 060707 | 3.43 | 22.11 | 0.35 | 22.11 | Swift | 21.00 | 0.20 | | -1.69 | min | | C15 |
| 060714 | 2.71 | 21.98 | 0.32 | 1.50 | Swift | 21.80 | 0.10 | | -0.97 | min | | C15 |
| 060719 | 1.53 | 22.45 | 0.10 | 0.10 | Swift | | | | | | |
| 060729 | 0.54 | 20.94 | 0.10 | 0.12 | Swift | | | | | | |
| 060801 | 1.13 | 21.36 | 0.38 | 21.36 | Swift | | | | | | |
| 060814 | 1.92 | 22.46 | 0.05 | 0.05 | A16 | | | | | | |
| 060904B | 0.70 | 21.45 | 0.22 | 0.37 | Swift | | | | | | |
| 060906 | 3.69 | 22.40 | 0.00 | 22.40 | Swift | 21.85 | 0.10 | | -1.72 | min | | C15 |
| 060908 | 1.88 | 21.99 | 0.19 | 0.29 | A16 | | | | | | |
| 060912A | 0.94 | 21.61 | 0.19 | 0.30 | A16 | | | | | | |
| 060926 | 3.21 | 22.70 | 0.26 | 0.40 | Swift | 22.60 | 0.15 | | -1.32 | min | | C15 |
| 060927 | 5.47 | 21.97 | 0.73 | 21.97 | A16 | 22.50 | 0.15 | | -1.55 | min | | C15 |

| GRB | | | | | | | | | | | |
|---|---|---|---|---|---|---|---|---|---|---|---|
| 061007 | 1.26 | 21.83 | 0.02 | 0.02 | A16 | | | -0.53 | 0.18 | 0.13 | V17 |
| 061021 | 0.35 | 20.87 | 0.13 | 0.19 | A16 | | | | | | |
| 061110A | 0.76 | 21.85 | 0.00 | 21.85 | Swift | | | | | | |
| 061110B | 3.44 | 22.38 | 0.43 | 22.38 | Swift | 22.35 | 0.10 | -1.84 | min | | C15 |
| 061121 | 1.31 | 21.87 | 0.07 | 0.08 | A16 | | | -0.19 | 0.09 | 0.06 | V17 |
| 061202 | 2.25 | 23.19 | 0.05 | 0.06 | Swift | | | | | | |
| 061222A | 2.09 | 22.72 | 0.06 | 0.06 | Swift | | | | | | |
| 061222B | 3.36 | 23.29 | 0.00 | 23.29 | Swift | | | | | | |
| 070110 | 2.35 | 21.53 | 0.21 | 0.39 | Swift | 21.70 | 0.10 | -1.32 | min | | C15 |
| 070125 | 1.55 | 21.60 | 0.30 | 21.60 | Swift | | | | | | |
| 070129 | 2.34 | 22.18 | 0.15 | 0.22 | Swift | | | | | | |
| 070208 | 1.17 | 22.00 | 0.15 | 0.15 | Swift | | | | | | |
| 070223 | 1.63 | 22.76 | 0.18 | 0.22 | Swift | | | | | | |
| 070306 | 1.50 | 22.56 | 0.06 | 0.06 | Swift | | | -0.24 | 0.80 | 0.80 | V17 |
| 070318 | 0.84 | 21.98 | 0.08 | 0.09 | Swift | | | | | | |
| 070328 | 2.06 | 22.45 | 0.03 | 0.03 | A16 | | | | | | |
| 070411 | 2.95 | 22.37 | 0.00 | 22.37 | Swift | 19.89 | 0.40 | IC | | | |
| 070419B | 1.96 | 21.89 | 0.14 | 0.19 | Swift | | | | | | |
| 070506 | 2.31 | 21.65 | 0.66 | 21.65 | Swift | 22.00 | 0.30 | -0.65 | min | | C15 |
| 070508 | 0.82 | 21.76 | 0.16 | 0.21 | Swift | | | | | | |
| 070521 | 2.09 | 23.19 | 0.06 | 0.06 | Swift | | | | | | |
| 070529 | 2.50 | 22.45 | 0.39 | 22.45 | Swift | | | | | | |
| 070611 | 2.04 | 21.88 | 0.00 | 21.88 | Swift | 21.30 | 0.20 | | | | |
| 070714B | 0.92 | 21.58 | 0.00 | 21.58 | Swift | | | | | | |
| 070721B | 3.63 | 22.11 | 0.26 | 0.61 | Swift | 21.50 | 0.20 | -2.14 | min | | C15 |
| 070802 | 2.45 | 22.20 | 0.36 | 1.20 | Swift | 21.50 | 0.20 | -0.54 | min | | C15 |
| 070810A | 2.17 | 22.00 | 0.20 | 0.30 | Swift | 21.70 | 0.20 | | | | |
| 071003 | 1.60 | 21.43 | 0.00 | 21.43 | Swift | | | | | | |
| 071010A | 0.98 | 22.31 | 0.21 | 0.26 | Swift | | | | | | |
| 071010B | 0.95 | 21.48 | 0.32 | 1.00 | Swift | | | | | | |
| 071020 | 2.15 | 21.80 | 0.14 | 0.20 | A16 | | | | | | |
| 071021 | 2.45 | 22.30 | 0.11 | 0.15 | Swift | | | | | | |
| 071025 | 4.88 | 22.54 | 0.16 | 0.24 | Swift | | | | | | |
| 071031 | 2.69 | 22.12 | 0.00 | 22.12 | Swift | 22.15 | 0.05 | -1.73 | 0.05 | 0.05 | A18 |
| 071112C | 0.82 | 21.09 | 0.22 | 0.42 | A16 | | | | | | |
| 071117 | 1.33 | 22.22 | 0.12 | 0.14 | A16 | | | -0.29 | 0.15 | 0.09 | V17 |
| 071122 | 1.14 | 21.26 | 0.29 | 0.95 | Swift | | | | | | |
| 071227 | 0.38 | 21.61 | 0.00 | 21.61 | Swift | 21.10 | 0.30 | | | | |
| 080205 | 2.72 | 22.51 | 0.31 | 0.55 | Swift | | | | | | |
| 080207 | 2.09 | 23.23 | 0.06 | 0.07 | Swift | | | | | | |
| 080210 | 2.64 | 22.32 | 0.21 | 0.32 | Swift | 21.90 | 0.10 | | | | |
| 080310 | 2.43 | 21.70 | 0.30 | 0.70 | Swift | 19.69 | 0.30 | IC | | | |
| 080319A | 2.03 | 20.11 | 2.20 | 20.11 | Swift | | | | | | |
| 080319B | 0.94 | 21.20 | 0.04 | 0.04 | A16 | | | | | | |
| 080319C | 1.95 | 22.00 | 0.14 | 0.18 | A16 | | | | | | |
| 080325 | 1.78 | 22.33 | 0.20 | 0.26 | Swift | | | | | | |
| 080330 | 1.51 | 21.72 | 0.20 | 0.34 | Swift | | | | | | |
| 080411 | 1.03 | 21.76 | 0.05 | 0.06 | Swift | | | -1.56 | min | | C15 |
| 080413A | 2.43 | 21.38 | 0.91 | 21.38 | Swift | 21.85 | 0.15 | | | | |
| 080413B | 1.10 | 21.46 | 0.14 | 0.18 | Swift | | | -0.29 | 0.20 | 0.20 | V17 |
| 080430 | 0.77 | 21.74 | 0.07 | 0.08 | Swift | | | | | | |
| 080520 | 1.55 | 22.70 | 0.38 | 0.70 | Swift | | | | | | |
| 080602 | 1.82 | 22.18 | 0.10 | 0.13 | Swift | | | -0.13 | 0.20 | 0.30 | V17 |
| 080603A | 1.69 | 22.00 | 0.20 | 0.30 | Swift | | | | | | |
| 080603B | 2.69 | 22.01 | 0.17 | 0.26 | A16 | 21.85 | 0.05 | | | | |
| 080604 | 1.42 | 21.46 | 0.63 | 1.46 | Swift | | | | | | |
| 080605 | 1.64 | 22.04 | 0.05 | 0.05 | A16 | | | -0.23 | 0.80 | 0.80 | V17 |
| 080607 | 3.04 | 22.56 | 0.09 | 0.11 | A16 | 22.70 | 0.15 | -0.20 | | | B19 |
| 080707 | 1.23 | 21.70 | 0.26 | 0.40 | Swift | | | | | | |
| 080710 | 0.85 | 21.15 | 0.24 | 0.54 | Swift | | | -1.73 | min | | C15 |
| 080721 | 2.59 | 22.00 | 0.04 | 0.04 | A16 | 21.60 | 0.10 | | | | |
| 080804 | 2.20 | 21.15 | 0.59 | 21.15 | Swift | 21.30 | 0.15 | -0.75 | min | | C15 |
| 080805 | 1.50 | 22.26 | 0.21 | 0.26 | Swift | | | | | | |
| 080810 | 3.36 | 21.60 | 0.44 | 21.60 | Swift | 19.70 | 0.40 | IC | | | |
| 080905A | 0.12 | 21.68 | 0.00 | 21.68 | Swift | | | | | | |
| 080905B | 2.37 | 22.57 | 0.10 | 0.12 | Swift | 22.60 | 0.30 | | | | |
| 080913 | 6.73 | 23.00 | 0.66 | 23.00 | Swift | 19.98 | 0.40 | IC | | | |
| 080916A | 0.69 | 21.91 | 0.11 | 0.12 | Swift | | | | | | |
| 080928 | 1.69 | 21.56 | 0.28 | 0.86 | Swift | | | | | | |
| 081008 | 1.97 | 21.41 | 0.46 | 21.41 | Swift | 21.59 | 0.10 | -0.52 | 0.11 | 0.11 | A18 |
| 081028 | 3.04 | 21.70 | 0.26 | 0.70 | Swift | | | | | | |
| 081029 | 3.85 | 21.48 | 0.60 | 21.48 | Swift | 21.45 | 0.10 | | | | |
| 081109 | 0.98 | 22.16 | 0.08 | 0.09 | Swift | | | | | | |
| 081118 | 2.58 | 22.00 | 0.41 | 22.00 | Swift | 21.50 | 0.20 | | | | |
| 081121 | 2.51 | 21.70 | 0.20 | 0.34 | A16 | | | | | | |
| 081203A | 2.05 | 21.90 | 0.14 | 0.20 | Swift | 22.00 | 0.10 | | | | |
| 081210 | 2.06 | 21.79 | 0.00 | 21.79 | Swift | | | | | | |
| 081221 | 2.26 | 22.53 | 0.07 | 0.07 | A16 | | | | | | |
| 081222 | 2.77 | 21.68 | 0.13 | 0.18 | A16 | 20.80 | 0.20 | | | | |
| 090102 | 1.55 | 21.91 | 0.11 | 0.13 | A16 | | | | | | |
| 090201 | 2.10 | 23.01 | 0.05 | 0.05 | Swift | | | | | | |
| 090205 | 4.65 | 22.23 | 0.53 | 22.23 | Swift | 20.73 | 0.05 | -0.57 | min | | C15 |
| 090313 | 3.38 | 22.64 | 0.13 | 0.18 | Swift | 21.30 | 0.20 | -1.40 | 0.30 | 0.30 | A18 |
| 090418A | 1.61 | 22.23 | 0.07 | 0.08 | Swift | | | | | | |
| 090323 | 3.58 | 22.34 | 0.00 | 22.34 | Swift | 20.75 | 0.10 | 0.25 | 0.09 | 0.09 | A18 |
| 090328A | 0.74 | 21.78 | 0.26 | 0.48 | Swift | | | | | | |
| 090404 | 3.00 | 23.09 | 0.06 | 0.07 | Swift | | | | | | |
| 090417B | 0.35 | 22.52 | 0.05 | 0.04 | Swift | | | | | | |

| GRB | | | | | | | | | | | |
|---|---|---|---|---|---|---|---|---|---|---|---|
| 090423 | 8.20 | 23.00 | 0.28 | 0.70 | Swift | | | | | | |
| 090424 | 0.54 | 21.82 | 0.06 | 0.06 | A16 | | | | | | |
| 090426 | 2.61 | 21.64 | 0.66 | 21.64 | Swift | 19.88 | 0.30 | IC | | | |
| 090429B | 9.40 | 23.15 | 0.03 | 0.03 | S15 | | | | | | |
| 090510 | 0.90 | 21.53 | 0.33 | 0.93 | Swift | | | | | | |
| 090516A | 4.11 | 22.38 | 0.14 | 0.18 | Swift | 21.73 | 0.10 | -1.36 | min | | C15 |
| 090519 | 3.85 | 22.97 | 0.00 | 22.97 | Swift | 21.00 | 0.40 | | | | |
| 090529 | 2.62 | 22.23 | 0.00 | 22.23 | Swift | 20.30 | 0.30 | | | | |
| 090530 | 1.27 | 21.74 | 0.16 | 0.24 | Swift | | | | | | |
| 090618 | 0.54 | 21.40 | 0.06 | 0.08 | Swift | | | | | | |
| 090709A | 1.80 | 22.32 | 0.08 | 0.07 | Swift | | | | | | |
| 090715B | 3.01 | 22.07 | 0.19 | 0.31 | A16 | 21.65 | 0.15 | | | | |
| 090726 | 2.71 | 22.18 | 0.20 | 0.33 | Swift | 21.80 | 0.30 | | | | |
| 090809 | 2.74 | 21.85 | 0.27 | 0.85 | Swift | 21.70 | 0.20 | -0.86 | 0.13 | 0.35 | B19 |
| 090812 | 2.45 | 22.26 | 0.22 | 0.34 | Swift | 22.30 | 0.10 | -1.64 | min | | C15 |
| 090814A | 0.70 | 21.34 | 0.50 | 21.34 | Swift | | | | | | |
| 090902B | 1.82 | 22.34 | 0.12 | 0.14 | Swift | | | | | | |
| 090926A | 2.11 | 21.74 | 0.19 | 0.17 | Z18 | 21.55 | 0.10 | -1.97 | 0.11 | 0.02 | B19 |
| 090926B | 1.24 | 22.34 | 0.05 | 0.05 | A16 | | | -0.25 | 0.18 | 0.20 | V17 |
| 090927 | 1.37 | 21.54 | 0.29 | 0.77 | Swift | | | | | | |
| 091003 | 0.90 | 21.53 | 0.19 | 0.30 | Swift | | | | | | |
| 091018 | 0.97 | 21.48 | 0.15 | 0.20 | Swift | | | | | | |
| 091020 | 1.71 | 21.89 | 0.09 | 0.11 | Swift | | | | | | |
| 091024 | 1.09 | 22.79 | 0.16 | 0.22 | Swift | | | | | | |
| 091029 | 2.75 | 21.95 | 0.16 | 0.26 | Swift | 20.70 | 0.15 | | | | |
| 091109A | 3.08 | 22.26 | 0.29 | 0.65 | Swift | | | | | | |
| 091127 | 0.49 | 21.11 | 0.16 | 0.27 | Swift | | | | | | |
| 091208B | 1.06 | 22.02 | 0.12 | 0.15 | Swift | | | | | | |
| 100117A | 0.92 | 21.61 | 0.16 | 0.21 | Swift | | | | | | |
| 100219A | 4.67 | 22.70 | 0.34 | 22.70 | Swift | 21.20 | 0.20 | -1.24 | 0.05 | 0.07 | B19 |
| 100302A | 4.81 | 22.26 | 0.46 | 22.26 | Swift | 20.50 | 0.30 | | | | |
| 100316A | 3.16 | 20.82 | 0.50 | 20.82 | Swift | 22.20 | 0.25 | | | | |
| 100316B | 1.18 | 22.15 | 0.65 | 22.15 | Swift | | | | | | |
| 100316D | 0.06 | 21.71 | 0.63 | 21.71 | Swift | | | | | | |
| 100418A | 0.62 | 21.45 | 0.22 | 0.24 | z18 | | | | | | |
| 100424A | 2.47 | 22.78 | 0.88 | 22.78 | Swift | | | | | | |
| 100425A | 1.76 | 21.77 | 0.00 | 21.77 | Swift | 21.00 | 0.20 | -0.96 | 0.42 | 0.42 | C15 |
| 100513A | 4.77 | 22.78 | 0.26 | 0.48 | Swift | 21.80 | 0.05 | | | | |
| 100615A | 1.40 | 23.26 | 0.06 | 0.06 | Swift | | | -0.55 | 0.26 | 0.22 | V17 |
| 100621A | 0.54 | 22.38 | 0.03 | 0.02 | A16 | | | | | | |
| 100724A | 1.29 | 21.15 | 0.64 | 21.15 | Swift | | | | | | |
| 100728A | 1.57 | 22.38 | 0.07 | 0.08 | Swift | | | | | | |
| 100728B | 2.11 | 21.85 | 0.30 | 0.85 | Swift | 21.20 | 0.50 | | | | |
| 100814A | 1.44 | 21.18 | 0.17 | 0.27 | Swift | | | | | | |
| 100816A | 0.81 | 21.62 | 0.22 | 0.34 | Swift | | | | | | |
| 100901A | 1.41 | 21.34 | 0.19 | 0.26 | z18 | | | | | | |
| 100906A | 1.73 | 21.78 | 0.30 | 0.78 | Swift | | | | | | |
| 101219A | 0.72 | 22.28 | 0.62 | 0.98 | Swift | | | | | | |
| 101219B | 0.55 | 20.90 | 0.21 | 0.20 | z18 | | | | | | |
| 101225A | 0.85 | 21.18 | 0.15 | 0.18 | Swift | | | | | | |
| 110128A | 2.34 | 22.12 | 0.00 | 22.12 | Swift | 21.90 | 0.15 | | | | |
| 110205A | 2.22 | 21.71 | 0.20 | 0.33 | Swift | 21.45 | 0.20 | -0.82 | min | | C15 |
| 110213A | 1.46 | 20.72 | 0.76 | 20.72 | Swift | | | | | | |
| 110422A | 1.77 | 22.18 | 0.08 | 0.10 | Swift | | | | | | |
| 110503A | 1.61 | 21.22 | 0.19 | 0.32 | A16 | | | | | | |
| 110709B | 2.09 | 21.18 | 0.05 | 0.06 | A16 | | | | | | |
| 110715A | 0.82 | 22.20 | 0.12 | 0.12 | Swift | | | | | | |
| 110731A | 2.83 | 21.95 | 0.39 | 21.95 | Swift | 21.90 | 0.30 | | | | |
| 110801A | 1.89 | 21.59 | 0.24 | 0.51 | Swift | | | | | | |
| 110808A | 1.35 | 21.90 | 0.24 | 0.43 | Swift | | | -0.76 | 0.27 | 0.27 | A18 |
| 110818A | 3.36 | 22.18 | 0.32 | 1.18 | Swift | 21.90 | 0.40 | | | | |
| 110918A | 0.89 | 21.24 | 0.20 | 0.31 | Swift | | | 0.24 | 0.11 | 0.11 | A18 |
| 111008A | 4.99 | 22.33 | 0.19 | 0.29 | z18 | 22.40 | 0.10 | -1.48 | 0.31 | 0.31 | B19 |
| 111107A | 2.89 | 21.86 | 0.63 | 21.86 | Swift | 21.00 | 0.20 | -0.74 | 0.20 | 0.20 | B19 |
| 111201A | 3.39 | 22.11 | 0.43 | 22.11 | Swift | | | | | | |
| 111209A | 0.68 | 21.36 | 0.09 | 0.11 | Swift | | | | | | |
| 111225A | 0.30 | 21.48 | 0.48 | 21.48 | Swift | | | | | | |
| 111228A | 0.71 | 21.59 | 0.08 | 0.10 | Swift | | | | | | |
| 111229A | 1.38 | 21.60 | 0.30 | 0.60 | Swift | | | | | | |
| 120118B | 2.94 | 22.95 | 0.16 | 0.18 | Swift | | | -0.80 | 0.23 | 0.17 | A18 |
| 120119A | 1.73 | 21.97 | 0.23 | 0.26 | z18 | 22.60 | 0.20 | -0.09 | 0.14 | 0.14 | A18 |
| 120326A | 1.80 | 21.85 | 0.10 | 0.12 | A16 | | | | | | |
| 120327A | 2.81 | 21.69 | 0.71 | 21.69 | Swift | 22.00 | 0.05 | -1.49 | 0.04 | 0.04 | B19 |
| 120404A | 2.88 | 21.90 | 0.33 | 21.90 | Swift | 20.70 | 0.30 | -0.30 | 0.09 | 0.09 | A18 |
| 120422A | 0.28 | 21.25 | 0.00 | 21.25 | Swift | | | | | | |
| 120521C | 6.00 | 22.85 | 0.33 | 22.85 | Swift | | | | | | |
| 120711A | 1.41 | 22.32 | 0.09 | 0.09 | Swift | | | | | | |
| 120712A | 4.17 | 22.38 | 0.26 | 0.60 | Swift | 19.95 | 0.15 | -0.38 | max | max | B19 |
| 120714B | 0.40 | 21.51 | 0.00 | 21.51 | Swift | | | -0.30 | 0.11 | 0.11 | A18 |
| 120716A | 2.49 | 22.73 | 0.63 | 22.73 | Swift | 22.00 | 0.15 | -0.71 | 0.06 | 0.06 | B19 |
| 120724A | 1.48 | 21.48 | 0.56 | 21.48 | Swift | | | | | | |
| 120729A | 0.80 | 21.51 | 0.21 | 0.21 | A16 | | | | | | |
| 120802A | 3.80 | 22.33 | 0.32 | 1.23 | A16 | | | | | | |
| 120811C | 2.67 | 22.26 | 0.20 | 0.31 | Swift | 21.50 | 0.15 | | | | |
| 120815A | 2.36 | 21.48 | 0.56 | 21.48 | Swift | 22.05 | 0.10 | -1.45 | 0.03 | 0.03 | B19 |
| 120907A | 0.97 | 21.20 | 0.27 | 0.73 | Swift | | | | | | |
| 120909A | 3.93 | 22.41 | 0.16 | 0.24 | Swift | 21.70 | 0.10 | -1.06 | 0.20 | 0.20 | B19 |
| 120923A | 7.84 | 23.88 | 0.00 | 23.88 | Swift | | | | | | |
| 121024A | 2.30 | 22.08 | 0.20 | 0.30 | Swift | 21.85 | 0.10 | -0.76 | 0.06 | 0.06 | B19 |
| 121027A | 1.77 | 22.30 | 0.06 | 0.07 | Swift | 22.80 | 0.30 | | | | |
| 121123A | 2.70 | 21.95 | 0.20 | 0.34 | A16 | | | | | | |
| 121128A | 2.20 | 21.80 | 0.60 | 21.80 | Swift | 21.80 | 0.25 | | | | |
| 121201A | 3.39 | 22.11 | 0.43 | 22.11 | Swift | 22.00 | 0.20 | | | | |
| 121209A | 2.10 | 23.06 | 0.06 | 0.07 | Swift | | | | | | |
| 121211A | 1.02 | 21.89 | 0.09 | 0.09 | Swift | | | | | | |
| 121229A | 2.71 | 22.49 | 0.00 | 22.49 | Swift | 21.70 | 0.20 | | | | |
| 130408A | 3.76 | 21.94 | 0.00 | 21.94 | Swift | 21.80 | 0.10 | -1.48 | 0.07 | 0.07 | B19 |
| 130418A | 1.22 | 20.88 | 0.69 | 20.88 | Swift | | | | | | |
| 130420A | 1.30 | 21.66 | 0.11 | 0.14 | A16 | | | | | | |
| 130427A | 0.34 | 20.90 | 0.10 | 0.12 | Z18 | 21.90 | 0.30 | | | | |
| 130427B | 2.78 | 21.85 | 0.42 | 21.85 | A16 | | | | | | |
| 130505A | 2.27 | 21.95 | 0.09 | 0.11 | A16 | 20.65 | 0.10 | -1.42 | min | | C15 |
| 130511A | 1.30 | 21.97 | 0.33 | 0.76 | Swift | | | | | | |
| `130518A | 2.49 | | | | | 21.80 | 0.20 | | | | |
| 130603B | 0.36 | 21.66 | 0.09 | 0.10 | Swift | | | | | | |
| 130606A | 5.91 | 22.43 | 0.34 | 22.43 | Swift | 19.91 | 0.02 | -1.83 | 0.10 | 0.10 | B19 |
| 130610A | 2.09 | 21.48 | 0.48 | 21.48 | Swift | 21.30 | 0.20 | | | | |
| 130612A | 2.01 | 22.15 | 0.36 | 1.15 | Swift | 22.10 | 0.30 | | | | |
| 130701A | 1.16 | 21.42 | 0.26 | 0.62 | A16 | | | | | | |
| 130702A | 0.15 | 20.78 | 0.22 | 0.48 | Swift | | | | | | |
| 130831A | 0.48 | 20.85 | 0.33 | 20.85 | A16 | | | | | | |
| 130907A | 1.24 | 22.04 | 0.03 | 0.04 | A16 | | | | | | |
| 130925A | 0.35 | 22.29 | 0.02 | 0.02 | LJ15 | | | 0.04 | 0.08 | 0.08 | A18 |
| 131004A | 0.72 | 21.48 | 0.48 | 21.48 | Swift | | | | | | |
| 131011A | 1.87 | | | | | 22.00 | 0.30 | | | | |
| 131030A | 1.29 | 21.65 | 0.13 | 0.18 | A16 | | | | | | |
| 131103A | 0.90 | 22.20 | 0.14 | 0.16 | Swift | | | | | | |
| 131105A | 1.69 | 22.21 | 0.17 | 0.23 | A16 | | | | | | |
| 131108A | 2.40 | 21.90 | 0.24 | 0.43 | Swift | 20.95 | 0.15 | | | | |
| 131117A | 4.04 | 22.46 | 0.35 | 22.46 | Swift | 20.00 | 0.30 | | | | |
| 131231A | 0.64 | 21.39 | 0.15 | 0.22 | Swift | | | -0.24 | 0.11 | 0.11 | A18 |
| 140206A | 2.73 | 22.31 | 0.08 | 0.09 | A16 | 21.50 | 0.20 | | | | |
| 140213A | 1.21 | 21.08 | 0.30 | 21.08 | Swift | | | | | | |
| 140215A | 0.00 | 21.15 | 0.23 | 0.42 | A16 | | | | | | |
| 140226A | 1.97 | 20.30 | 1.72 | 20.30 | Swift | 20.60 | 0.20 | -0.54 | min | | C15 |
| 140301A | 1.42 | 22.18 | 0.23 | 0.32 | Swift | | | | | | |
| 140304A | 5.28 | 22.70 | 0.38 | 22.70 | Swift | 21.60 | | -1.65 | min | | C15 |
| 140311A | 4.95 | 22.47 | 0.00 | 22.47 | Swift | 22.40 | 0.15 | -1.65 | 0.14 | 0.14 | B19 |
| 140318A | 1.02 | 21.90 | 0.28 | 0.76 | LJ15 | | | | | | |
| 140419A | 3.96 | 22.04 | 0.19 | 0.34 | Swift | 19.69 | 0.40 | IC | | | |
| 140423A | 3.26 | 22.08 | 0.15 | 0.23 | Swift | 20.45 | 0.20 | -1.44 | min | | C15 |
| 140430A | 1.60 | 21.95 | 0.22 | 0.35 | Swift | 21.80 | 0.30 | -0.02 | 0.19 | 0.19 | A18 |
| 140506A | 0.89 | 21.94 | 0.07 | 0.08 | A16 | | | | | | |
| 140508A | 1.03 | 21.41 | 0.28 | 0.72 | Swift | | | | | | |
| 140512A | 0.73 | 21.51 | 0.08 | 0.09 | A16 | | | | | | |
| 140515A | 6.32 | 22.60 | 0.30 | 0.60 | Swift | 19.70 | 0.40 | IC | | | |
| 140518A | 4.71 | 22.48 | 0.37 | 22.48 | Swift | 21.65 | 0.10 | | | | |
| 140614A | 4.23 | 22.40 | 0.39 | 22.40 | Swift | 21.60 | 0.30 | | | | |
| 140622A | 0.96 | 21.30 | 0.00 | 21.30 | Swift | | | | | | |
| 140629A | 2.28 | 21.75 | 0.22 | 0.41 | A16 | 22.00 | 0.30 | | | | |
| 140703A | 3.14 | 22.15 | 0.19 | 0.32 | A16 | 21.90 | 0.10 | | | | |
| 140710A | 0.56 | 21.51 | 0.00 | 21.51 | LJ15 | | | | | | |
| 140808A | 3.29 | 22.48 | 0.60 | 22.48 | Swift | 21.30 | 0.20 | | | | |
| 140903A | 0.35 | 21.26 | 0.19 | 0.30 | Swift | | | | | | |
| 140907A | 1.21 | 21.85 | 0.23 | 0.37 | Swift | | | | | | |
| 141028A | 2.33 | 20.92 | 0.60 | 20.92 | Swift | 20.60 | 0.15 | | -1.64 | 0.13 | 0.13 | B19 |
| 141109A | 2.99 | 22.34 | 0.13 | 0.17 | Swift | 22.10 | 0.10 | | -1.63 | 0.06 | 0.06 | B19 |
| 141121A | 1.47 | 21.62 | 0.18 | 0.30 | Swift | | | | | | |
| 141220A | 1.32 | 21.78 | 0.30 | 0.48 | Swift | | | | | | |
| 150101B | 0.09 | 20.56 | 0.58 | 20.56 | Swift | | | | | | |
| 150206A | 2.09 | 22.03 | 0.08 | 0.09 | Swift | 21.70 | 0.40 | | | | |
| 150301B | 1.52 | 21.32 | 0.72 | 21.32 | Swift | | | | | | |
| 150314A | 1.76 | 22.26 | 0.09 | 0.08 | Swift | | | | | | |
| 150403A | 2.06 | 21.90 | 0.08 | 0.10 | Swift | 21.80 | 0.20 | | -1.04 | 0.04 | 0.04 | B19 |
| 150413A | 3.14 | | | | | 22.10 | 0.20 | | | | |
| 150727A | 0.31 | 21.06 | 0.00 | 21.06 | Swift | | | | | | |
| 150821A | 0.76 | 22.34 | 0.09 | 0.09 | Swift | | | | | | |
| 150910A | 1.36 | 21.11 | 0.30 | 1.11 | Swift | | | | | | |
| 150915A | 1.97 | 22.36 | 0.98 | 1.36 | Swift | 21.20 | 0.30 | | | | |
| 151021A | 2.33 | 22.34 | 0.09 | 0.09 | Swift | 22.20 | 0.20 | | -0.98 | 0.07 | 0.07 | B19 |
| 151027A | 0.81 | 21.64 | 0.06 | 0.06 | Swift | | | | -0.76 | 0.17 | 0.17 | B19 |
| 151027B | 4.06 | 22.65 | 0.21 | 0.33 | Swift | 20.50 | 0.20 | | | | |
| 151029A | 1.42 | 22.11 | 0.31 | 0.56 | Swift | | | | | | |
| 151031A | 1.17 | 22.00 | 0.23 | 0.30 | Swift | | | | | | |
| 151215A | 2.59 | 22.18 | 0.92 | 22.18 | Swift | 21.30 | 0.30 | | | | |
| 160117B | 0.87 | 21.16 | 0.00 | 21.16 | Swift | | | | | | |
| 160131A | 0.97 | 21.61 | 0.09 | 0.09 | Swift | | | | | | |
| 160203A | 3.52 | 22.38 | 0.00 | 22.38 | Swift | 21.75 | 0.10 | | -1.31 | 0.04 | 0.04 | B19 |
| 160227A | 2.38 | 21.60 | 0.30 | 0.60 | Swift | 22.40 | 0.30 | | | | |
| 160314A | 0.73 | 21.00 | 0.00 | 21.00 | Swift | | | | | | |
| 160410A | 1.72 | 22.46 | 0.00 | 22.46 | Swift | | | | | | |
| 160425A | 0.56 | 22.00 | 0.10 | 0.10 | Swift | | | | | | |
| 160623A | 0.37 | 22.38 | 0.08 | 0.10 | Swift | | | | | | |
| 160625B | 1.41 | 20.46 | 1.21 | 20.46 | Swift | | | | | | |
| 160629A | 3.33 | 22.47 | 0.00 | 22.47 | Swift | 21.95 | 0.25 | | | | |
| 160804A | 0.74 | 21.56 | 0.13 | 0.18 | Swift | | | | | | |
| 161014A | 2.82 | 22.12 | 0.25 | 0.47 | Swift | 21.40 | 0.30 | | | | |
| 161017A | 2.01 | 21.32 | 0.29 | 1.02 | Swift | 20.50 | 0.30 | | | | |
| 161023A | 2.71 | 20.90 | 1.13 | 20.90 | Swift | 20.96 | 0.05 | -1.23 | 0.03 | 0.03 | B19 |
| 161219B | 0.15 | 21.31 | 0.06 | 0.06 | Swift | | | | | | |
| 170113A | 1.97 | 21.85 | 0.20 | 0.24 | Swift | | | | | | |
| 170202A | 3.65 | 21.45 | 0.78 | 21.45 | Swift | 21.55 | 0.10 | -1.28 | 0.09 | 0.09 | B19 |
| 170405A | 3.51 | 21.32 | 0.82 | 21.32 | Swift | 21.70 | 0.20 | | | | |
| 170519A | 0.82 | 21.65 | 0.09 | 0.11 | Swift | | | | | | |
| 170531B | 2.37 | 22.26 | 0.00 | 22.26 | Swift | 20.00 | 0.40 | | | | |
| 170604A | 1.33 | 20.97 | 0.39 | 20.97 | Swift | | | | | | |
| 170705A | 2.01 | 22.06 | 0.07 | 0.08 | Swift | | | | | | |
| 171020A | 1.87 | 22.26 | 0.30 | 0.65 | Swift | | | | | | |
| 171205A | 0.04 | 20.43 | 0.49 | 20.43 | Swift | | | | | | |
| 171222A | 2.41 | 22.17 | 0.26 | 0.46 | Swift | | | | | | |
| 180115A | 2.49 | 21.70 | 0.26 | 0.40 | Swift | 20.30 | 0.30 | | | | |
| 180205A | 1.41 | 21.08 | 0.66 | 21.08 | Swift | | | | | | |
| 180314A | 1.45 | 20.20 | 1.48 | 20.20 | Swift | | | | | | |
| 180325A | 2.04 | 22.15 | 0.11 | 0.15 | Swift | 22.30 | 0.14 | -0.96 | Min | | H19 |
| 180329B | 2.00 | 21.60 | 0.20 | 0.32 | Swift | 21.90 | 0.20 | | | | |
| 180404A | 1.00 | 21.76 | 0.00 | 21.76 | Swift | | | | | | |
| 180620B | 1.12 | 21.79 | 0.08 | 0.09 | Swift | | | | | | |
| 180624A | 2.86 | 22.00 | 0.20 | 0.40 | Swift | 22.50 | 0.20 | | | | |
| 180728A | 0.12 | 20.29 | 0.00 | 20.29 | Swift | | | | | | |
| 181010A | 1.39 | 22.28 | 0.08 | 0.07 | Swift | | | | | | |
| 181020A | 2.94 | 21.63 | 0.17 | 0.29 | Swift | 22.20 | 0.10 | -1.57 | 0.06 | 0.06 | H19 |
| 181110A | 1.51 | 20.78 | 0.68 | 20.78 | Swift | | | | | | |
| 181201A | 0.45 | 19.70 | 0.76 | 19.70 | Swift | | | | | | |
| 190106A | 1.86 | 20.92 | 0.53 | 20.92 | Swift | | | | | | |
| 190114A | 3.38 | 21.60 | 0.54 | 21.60 | Swift | | | -1.23 | 0.07 | 0.07 | H19 |
| 190114C | 0.42 | 22.89 | 0.03 | 0.04 | Swift | | | | | | |
| 190627A | 1.94 | 21.43 | 0.00 | 21.43 | Swift | | | | | | |